\documentclass[aps,prl,showpacs,preprintnumbers,amsmath,amssymb,twocolumn,floatfix,footinbib]{revtex4-1}

\usepackage{amsmath,amssymb}
\usepackage{graphicx}
\usepackage{bm}
\usepackage{bbm}
\usepackage{color}
\usepackage{natbib}
\usepackage{hyperref}
\hypersetup{
  colorlinks,
  citecolor=magenta,
  linkcolor=blue,
  urlcolor=blue}

\newcommand{\gbf}[1]{\boldsymbol #1}
\newcommand{\I}{\mathrm{i}}
\newcommand{\E}{\mathrm{e}}

\newcommand{\cfig}[1]{Fig.~\ref{#1}}
\newcommand{\ceqn}[1]{Eq.~(\ref{#1})}

\newcommand{\ctab}[1]{Tab.~\ref{#1}}
\newcommand{\pmin}{\phantom{-}}

\begin{document}
\title{Quantum Monte Carlo simulation of the chiral Heisenberg\\Gross-Neveu-Yukawa phase transition with a single Dirac cone}

\author{Thomas C. Lang}
\email{thomas.lang@uibk.ac.at}
\affiliation{Institute for Theoretical Physics, University of Innsbruck, 6020 Innsbruck, Austria}
\author{Andreas M. L\"{a}uchli}
\affiliation{Institute for Theoretical Physics, University of Innsbruck, 6020 Innsbruck, Austria}

\begin{abstract}
We present quantum Monte Carlo simulations for the chiral Heisenberg Gross-Neveu-Yukawa quantum phase transition of relativistic fermions with $N=4$ Dirac spinor components subject to a repulsive, local four fermion interaction in 2+1$d$. Here we employ a two dimensional lattice Hamiltonian with a single, spin-degenerate Dirac cone, which exactly reproduces a linear energy-momentum relation for all finite size lattice momenta in the absence of interactions. This allows us to significantly reduce finite size corrections compared to the widely studied honeycomb and $\pi$-flux lattices. A Hubbard term dynamically generates a mass beyond a critical coupling of ${U_c = 6.76(1)}$ as the system acquires antiferromagnetic order and SU(2) spin rotational symmetry is spontaneously broken. At the quantum phase transition we extract a self-consistent set of critical exponents ${\nu = 0.98(1)}$,  ${\eta_{\phi} = 0.53(1)}$, ${\eta_{\psi} = 0.18(1)}$, ${\beta = 0.75(1)}$. We provide evidence for the continuous degradation of the quasi-particle weight of the fermionic excitations as the critical point is approached from the semimetallic phase. Finally we study the effective "speed of light" of the low-energy relativistic description, which depends on the interaction $U$, but is expected to be regular across the quantum phase transition. We illustrate that the strongly coupled bosonic and fermionic excitations share a common velocity at the critical point.
\end{abstract}
\date{\today}


\maketitle

Right at the interface between bosonic spin and fermionic physics lies the Gross Neveu Yukawa (GNY) field theory, which is believed to capture the complex interplay of bosonic and fermionic (quantum) critical fluctuation giving rise to a large set of universal critical exponents \cite{Gross74,ZinnJustin91}. The universality class comprises the critical properties at the transition from a relativistic semi-metal described by massless Dirac fermions, to a symmetry broken phase with massive fermionic excitations, in which the order is captured by a $\mathbb{Z}_2$ (Ising), O($N$), or SU($N$) symmetric order parameters. In recent years it has become evident, that this physics of relativistic fermions is far from confined to high energy physics, but  manifests in many two and three dimensional condensed matter systems \cite{Vafek14,Wehling14,Armitage18,Young12,Young15}. Where the chiral Ising GNY transition \cite{Rosenstein93,Chandrasekharan13,Hands16,Hands93,Karkkainen94,Janssen14,Wang14,Wang15,Li15a,Li15b,Hesselmann16,Huffman17,Ihrig18,Zerf17} and the chiral-XY GNY transition \cite{Rosenstein93,Drut09,Chandrasekharan13,Hands16,Hands08,Chandrasekharan12,Scherer16,Jian17,Jiang17,Li17a,Li17b,Classen17,Wellegehausen17,Zerf17,Otsuka18,Xu18,Torres18} have been investigated extensively, far fewer results exist for the much more challenging chiral Heisenberg GNY transition in 2+1$d$ \cite{Rosenstein93,Herbut06,Herbut09a,Assaad13,Toldin15,Otsuka16,Zerf17,Gracey18,Knorr18,Buividovich18}, which we focus on in this manuscript.

\begin{figure}[tp]
  \centering
  \includegraphics[width=\columnwidth]{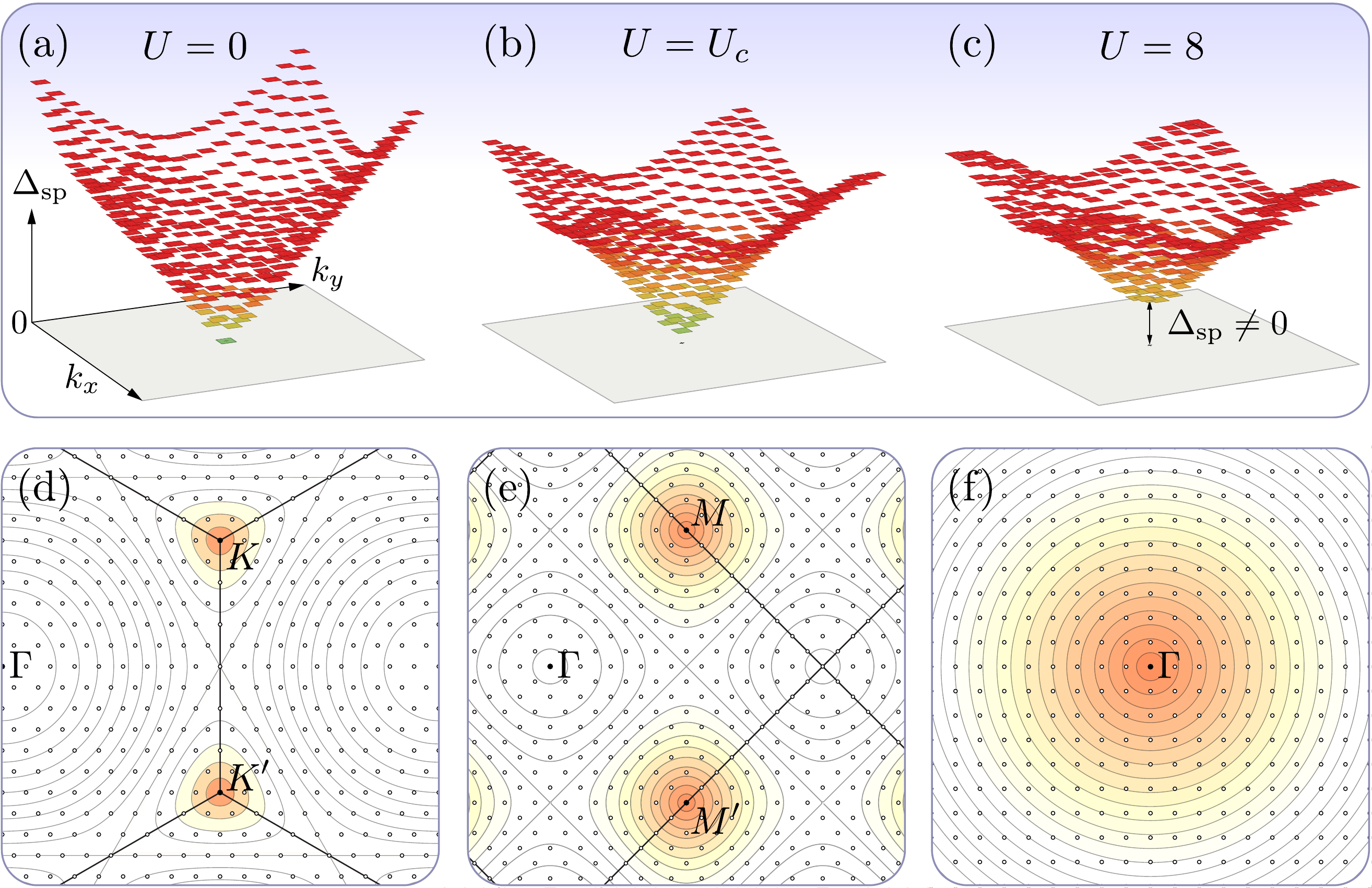}
  \caption{The momentum resolved single particle gap from QMC simulations (a) in the chiral limit (b) at the critical point and (c) in the massive phase in the first Brillouin zone for an $L=19$ system. The dispersion illustrates the interaction induced spontaneous mass generation for sufficiently strong interactions and the renormalization of the bandwidth across the quantum phase transition. To illustrate the portion of momenta with a linear dispersion close to the Dirac point (shaded areas) the finite size momentum resolution for an $L=18$ (d) honeycomb lattice (e) $\pi$-flux lattice (staggered fermions) and for (f) SLAC fermions is superimposed on the lines of constant energy in the Brillouin zone.}
  \label{fig:sp_gap}
\end{figure}

The issue shared among all the investigations is the apparent disparity between the estimates for the critical exponents not only between complementary methods, but even within different Monte Carlo simulations. The latter could be attributed to the fact that only a small region of the Brillouin zone of common lattices, such as the honeycomb lattice and $\pi$-flux (staggered fermions), actually displays relativistic behavior at low energies (cf. \cfig{fig:sp_gap}(d)--(f)) and large finite size lattices are required in order to obtain a sufficiently high momentum resolution at low energies to guarantee asymptotic scaling at criticality. 

In this manuscript we set out to minimize the finite size effects by implementing a single Dirac cone on the lattice, which allows to maximize the portion with a relativistic dispersion in the Brillouin zone. In addition, rather than distributing the fermion species across different momenta, or reducing the Brillouin zone to patches in momentum space \cite{Liu18}, a single Dirac cone is the closest representation of the continuum Dirac operator \cite{Susskind77,Wilson77,Nason85,Chandrasekharan04,Gattringer10}. Furthermore, a single spin degenerate ($N_f=2$) cone Dirac constitutes the smallest possible number of fermions species $N=2N_f=4$, or components of the Dirac spinor representation, for which an SU(2) symmetric order parameter can be formulated on a lattice. As such, our investigation provides a  benchmark for complementary approaches such as $\epsilon$- and $1/N$-expansions, where for small fermion species numbers their estimates for the critical exponents vary the most.

\paragraph{Model and Hamiltonian ---}
Here we consider a Hamiltonian formulation of relativistic massless fermions with a perfect Dirac cone in energy-momentum space in 2+1$d$. For each fermion flavor ${\sigma\in\{1,\ldots,N_f\}}$ the free Hamiltonian for a single Dirac cone reads ${H_{\mathbf{k}\sigma} = v_\text{F}^0 \sum_{\mathbf{k}}\gbf{\psi}^{\dagger}_{\mathbf{k}\sigma}\,\gbf{\sigma}\cdot\mathbf{k}\,\gbf{\psi}_{\mathbf{k}\sigma}}$, with the spinor ${\gbf{\psi}^{\dagger}_{\sigma}=(a^{\dagger}_{\sigma},b^{\dagger}_{\sigma})}$ and the vector of Pauli matrices ${\gbf{\sigma}=(\gbf{\sigma}_x, \gbf{\sigma}_y)}$. The corresponding single-particle spectrum is given by ${\gbf{\varepsilon}_{\pm}(\mathbf{k})} = \pm v_{{\rm F}}^0 |\mathbf{k}|$ with a $2N_f$-fold degeneracy at ${\mathbf{k}=(0,0)}$. On a square lattice with the primitive vectors in $x$- and $y$-direction, unit lattice constant, and the Fourier transform ${a_{\mathbf{r}\sigma} = \sum_{\mathbf{k}} \E^{-\I\mathbf{k}\cdot\mathbf{r}} a_{\mathbf{k}\sigma}/L}$, the Hamiltonian takes the form
\begin{eqnarray}
   H_{t\sigma} &=& -v_{{\rm F}}^0 \sum_{i=1}^{L^2} \left[\I \sum_{x=-L/2}^{L/2} t(x) \left(a_{i\sigma}^{\dagger} b_{i+x,\sigma}^{\phantom{\dagger}} - b_{i+x,\sigma}^{\dagger} a_{i\sigma} ^{\phantom{\dagger}}\right)\right. \nonumber\\
       &&+ \left. \sum_{y=-L/2}^{L/2} t(y) \left(a_{i\sigma}^{\dagger} b_{i+y,\sigma}^{\phantom{\dagger}} + b_{i+y,\sigma}^{\dagger} a_{i\sigma}^{\phantom{\dagger}}\right)\right] \;.
    \label{Ht}
\end{eqnarray}
Here $a_{i\sigma}^{\dagger}$ ($b_{i\sigma}^{\dagger}$) creates an electron with flavor $\sigma$ in an orbital $a$ ($b$) of unit cell $i$, while $i+x$ denotes the unit cell in $x$-direction at a distance $|x|$. In the following we choose the Fermi velocity ${v_{{\rm F}}^0 = 1}$ as unit of energy. This setup may be interpreted as a square lattice bilayer with ${2L^2}$ sites, where we have bipartitioned the lattice, such that all sites within a layer belong to the same orbital (sublattice) and we have bipartite interlayer hopping only. 

The discrete inverse Fourier transform of the Dirac operator yields the finite size hopping amplitudes ${t(r) = (-1)^r \pi/[L\sin(r\pi/L)]}$, ${r\ne 0}$, which in the thermodynamic limit (TDL) ${L\to\infty}$ implies that the hopping amplitude decays as ${t(r) = (-1)^r/r}$. Note, that a truncation of the hopping range introduces unwanted low energy states (doublers) \cite{Sugihara03}.
\begin{figure}[t]
  \centering
  \includegraphics[width=\columnwidth]{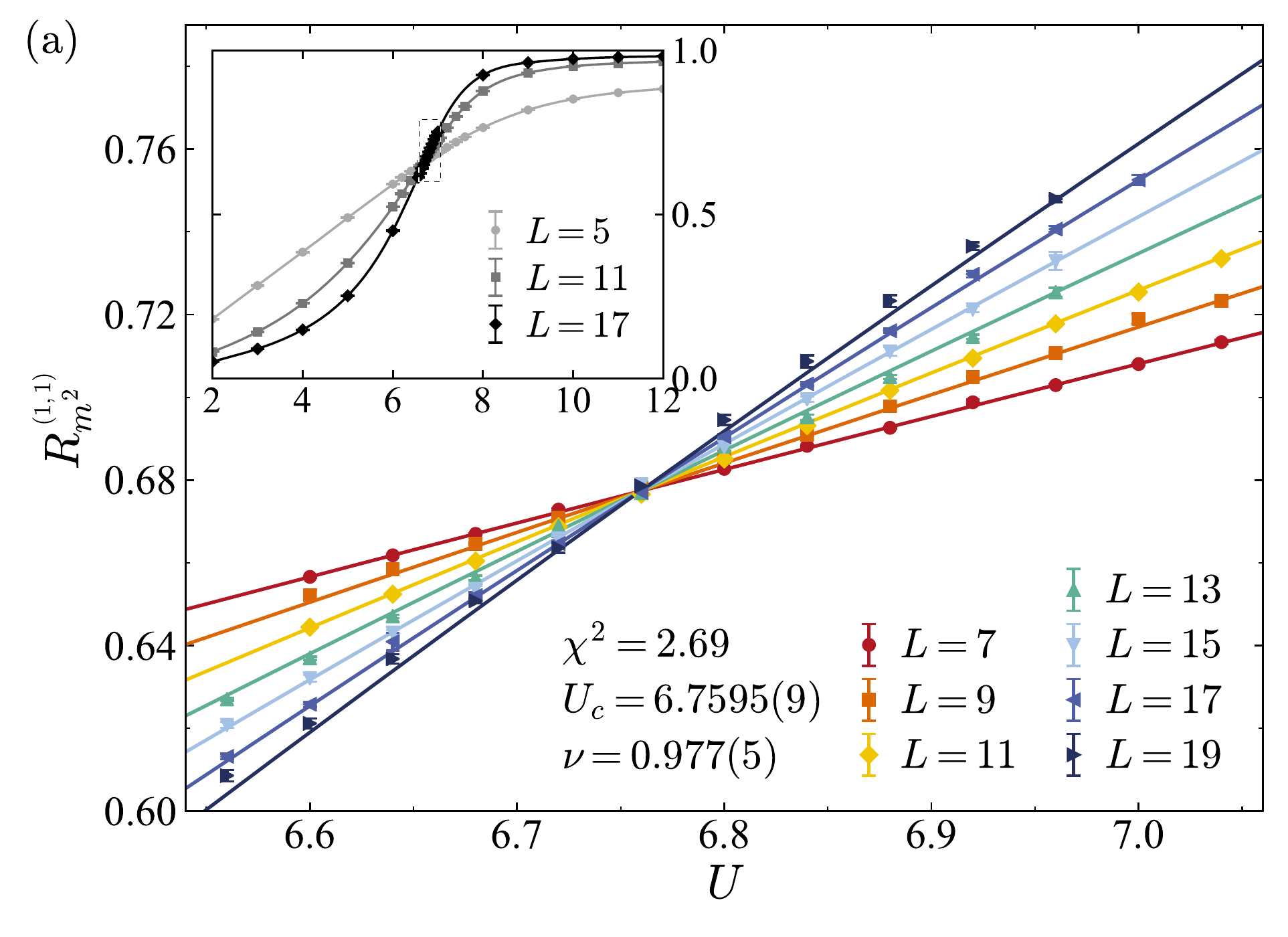}\\
  \includegraphics[width=\columnwidth]{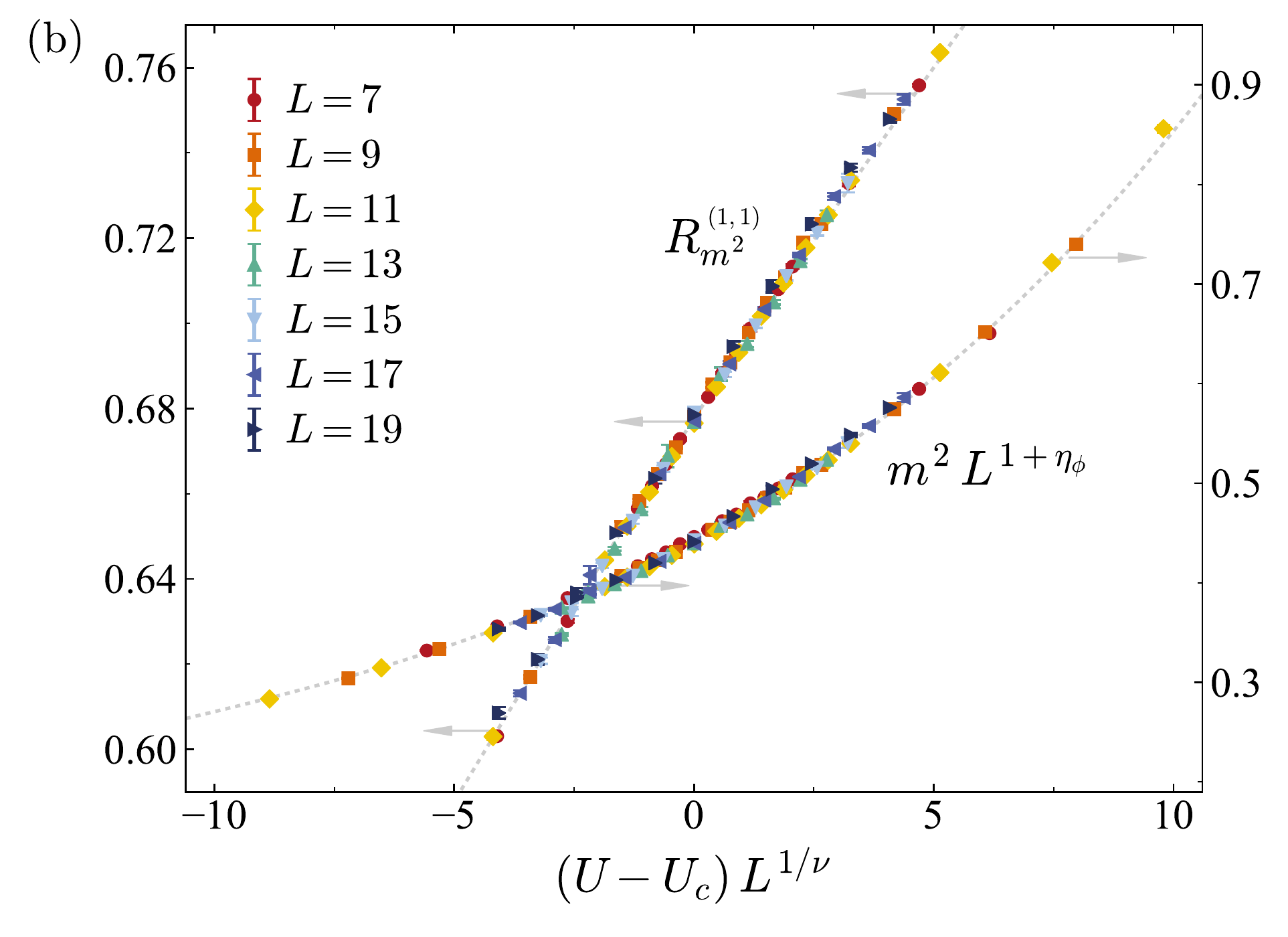}\vspace{-1em}
  \caption{The correlation ratio close to the critical point and for a larger coupling range (inset) (a) from which we extract the critical exponent $\nu$ from a fit to the data. (b) Data collapse of the correlation ratio (left scale) and the squared magnetization (right scale) using the critical exponents $\nu$ and $\eta_{\phi}$ extracted from the data in (a) and \cfig{fig:eta}(a), respectively. Dotted lines indicate the scaling functions.}
  \label{fig:nu}
\end{figure}
The lattice derivative in \ceqn{Ht} is the Hamiltonian formulation of SLAC fermions \cite{Drell76}. A variation, which corresponds to spinless fermions, has recently been used in Ref.~\cite{Li17b}. SLAC fermions avoid the Nielsen-Ninomiya theorem \cite{Nielsen81a,Nielsen81b,Nielsen81c} by violating locality on finite size lattices, but recover locality in the TDL for most of the Brillouin zone, but for the boundary \cite{Campos02}. The power-law hopping, which originates from the singularities of the engineered dispersion at the zone boundary, appears to be at odds with the locality condition of field theories and might raise concerns about their effect on the nature of the quantum phase transition. However, the hopping only runs along the major axes, and thus is not a genuine long-range coupling, as it does not couple a given site to a finite fraction of the total spatial volume.

In order to dynamically gap out the chiral fermions and to drive the system through a quantum phase transition we augment the free Hamiltonian with a local Hubbard-type repulsion ${H = \sum_{\sigma=1,2} H_{t\sigma} + \frac{U}{2} \sum_{i,c} \left(n_{i,c}-1\right)^2 }$, where ${n_{i,c} = \sum_{\sigma=1,2} c_{i\sigma}^{\dagger} c_{i\sigma}^{\phantom{\dagger}}}$ is the local density electrons in orbital $c\in \{a,b\}$. At strong coupling ${U \gg v_\text{F}^0}$ and half-filling, the Hamiltonian reduces to a bilayer Heisenberg model 
with antiferromagnetic Heisenberg interactions only between the layers. These interactions are not frustrated due to their bipartite structure, and we therefore expect N\'eel type antiferromagnetic long range order in this regime. The anticipated Dirac semi-metal to antiferromagnet (AFM) quantum phase transition is expected to be in the ${N=4}$ chiral Heisenberg GNY universality class. 

Finally the Hamiltonian $H_{t\sigma}$ is represented by a hermitian differentiation matrix and the Hubbard interaction can be decoupled at the cost of introducing a discrete auxiliary field via the Hubbard Stratonovich decomposition, which allows us to perform large-scale, sign-problem free auxiliary-field QMC simulations at zero temperature \cite{Sugiyama86,Assaad08,Li15b,Li16,Wei17}.

\begin{figure}[t]
  \centering
  \includegraphics[width=\columnwidth]{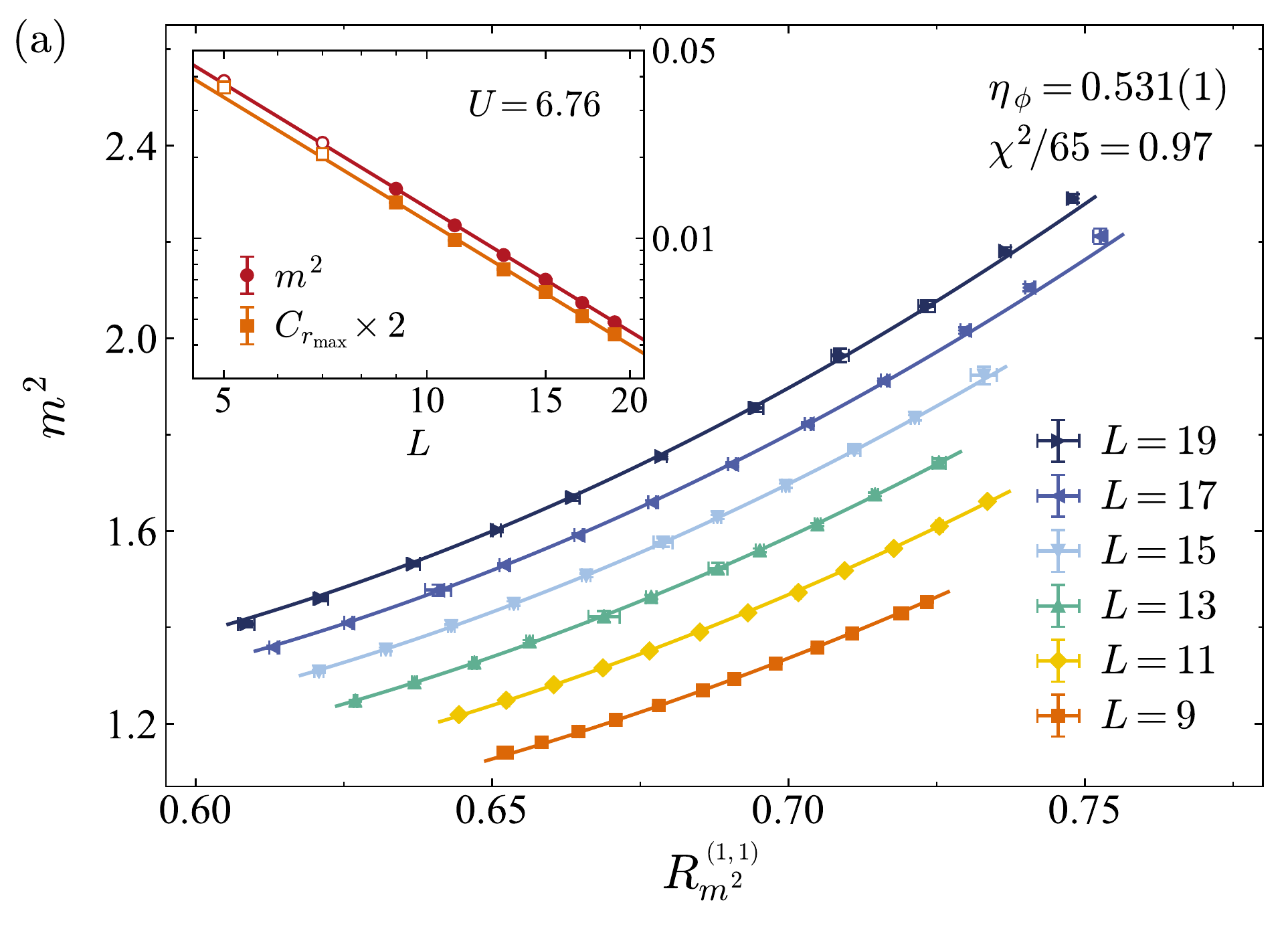}\\
  \includegraphics[width=\columnwidth]{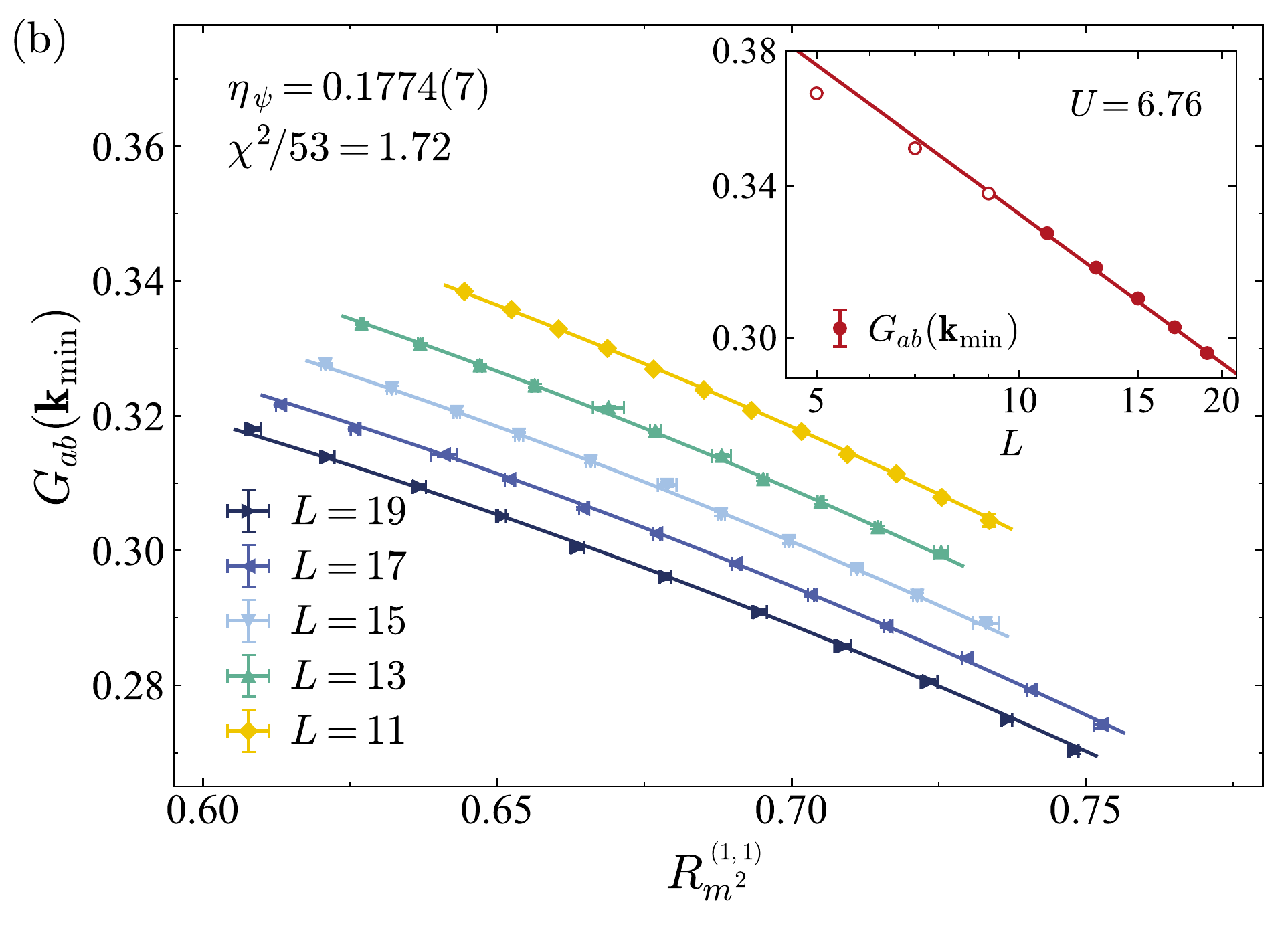}\vspace{-1em}
  \caption{Fit of the finite size scaling Ansatz (a) ${m^2(R_{m^2},L) = L^{1-\eta_{\phi}} f_0^m(R_{m^2})}$ to the squared magnetization (b) $G_{ab}(R_{m^2},L) = L^{-\eta_{\psi}} f_0^G(R_{m^2})$ to the off-diagonal component of the single particle correlation function in order to extract the bosonic and fermionic anomalous dimensions $\eta_{\phi}$ and $\eta_{\psi}$, respectively. The insets illustrate the compatibility of the estimated exponents with the finite size decay behavior of the correlations at the largest distance.}
  \label{fig:eta}
\end{figure}

\paragraph{QMC simulation results ---}
We track the emergence of long-range AFM order by measuring the spin structure factor ${S_\text{AFM}(\mathbf{k}) \equiv \sum_{\mathbf{r}}  \E^{\I\mathbf{k}\cdot\mathbf{r}} \langle\mathbf{S}(\mathbf{r})\cdot\mathbf{S}(\mathbf{0})\rangle/L^2}$, where ${\mathbf{S}(\mathbf{r}) = \mathbf{S}_{\mathbf{r}a}-\mathbf{S}_{\mathbf{r}b}}$ is the unit cell AFM order parameter with the spin ${\mathbf{S}_{\mathbf{r}a} = \frac{1}{2}a_{\mathbf{r}\alpha}^{\dagger}\gbf{\sigma}_{\alpha\beta} a_{\mathbf{r}\beta}}$ at position $\mathbf{r}$, orbital $a$, and $\gbf{\sigma}$ denotes the vector of the three Pauli matrices. In the bilayer setup AFM (N\'eel) order emerges at momentum ${\mathbf{Q} = (0,0)}$, such that ${S_\text{AFM}(\mathbf{Q})/L^2 = m^2}$. The evolution of the finite size magnetization as a function of the interaction strength is presented in the supplemental material (SM). Let us note that the stability of the semi-metal for small coupling $U<U_c$ is in agreement with the RG irrelevant interaction term, despite the aforementioned possibility of hopping induced long-range magnetic super exchange. The critical point, beyond which the fermions acquire mass can be precisely determined with the help of the RG invariant ratio ${R_{m^2}^{(n_1,n_2)} = 1 - S_\text{AFM}(\mathbf{Q} + n_1 \mathbf{b}_1 + n_2 \mathbf{b}_2)/S_\text{AFM}(\mathbf{Q})}$, where $\mathbf{b}_1$ and $\mathbf{b}_2$ denote the reciprocal lattice vectors \cite{Kaul15,Pujari16}. The correlation ratio scales to one (zero) in the ordered (disordered) phase, such that $R_{m^2}$ for different $L$ intersect at the critical coupling point and at a universal, but geometry dependent value $R_{m^2}^*$. Some variants of $R_{m^2}$ exhibit significantly less finite size corrections, such that we opt for $R_{m^2}^{(1,1)}$ with the least drift of the finite size crossing points (cf. SM). The correlation ratio data is shown in \cfig{fig:nu}(a) in the vicinity of the phase transition; the inset displays a wide range across the phase transition. We fit the data with the finite size scaling (FSS) Ansatz ${R_{m^2}(u,L) = f_0^R\!\left(u L^{1/\nu}\right) + L^{-\omega} f_1^R\!\left(u L^{1/\nu}\right)}$, where ${u=U-U_c}$ and we series expand the scaling functions $f_0^R$ and $f_1^R$ \cite{Campostrini14}. Adding corrections to scaling yields series expansion coefficients of $f_1^R$ and values for $\omega$, which vanish within error bars and  significantly degrades the quality of the data collapse. This behavior is in accordance with the lack of drift for larger lattices and reduces our scaling Ansatz to the first term without scaling corrections. For a fit to the data of systems ${L\ge 7}$ we obtain the critical coupling ${U = 6.759(1)}$ and the exponent associated with the correlation length ${\nu = 0.977(5)}$, which allows us to collapse the data in \cfig{fig:nu}(b).

Scaling the squared magnetization as a function of a dimensionless quantity, such as a correlation ratio $R_{m^2}$, allows us to eliminate the exponent $\nu$ from scaling and the FSS Ansatz reduces to ${m^2(R_{m^2},L) = L^{1-\eta_{\phi}} f_0^m (R_{m^2})}$, where we assume ${z=1}$ \cite{Campostrini14,Toldin15}. Figure~\ref{fig:eta}(a) shows the fit to the data, where we have expanded $f_0^m(R_{m^2})$ up to second order. We obtain a stable estimate $\eta_{\phi} = 0.531(1)$; Higher expansion orders do not alter the result and corrections to scaling behave similarly to the case for $\nu$ above. The inset of Fig.~\ref{fig:eta}(a) shows the compatibility of our estimate with a commonly used approach to extract $\eta_{\phi}$ from the scaling of the magnetization, or the spin correlations at the maximum distance $C(r_\text{max})$, which decay proportional to $L^{-(1+\eta_{\phi})}$ at the critical point. This would otherwise require a precise knowledge of $U_c$, since small deviations significantly alter the estimate for the critical exponent and would further involve fewer data points in the fit. The inset shows the compatibility of the exponent extracted in the main panel with the decay of the correlations function close to criticality. The solid symbols indicate the data points used to fit the amplitude of the decaying correlation function. To check the consistency of the extracted exponents we compute the critical exponent of the order parameter ${\beta = (1+\eta_{\phi})\nu/2 = 0.748(4)}$ and successfully perform a data collapse of the squared magnetization in \cfig{fig:nu}(b). In addition we compare with the data collapse obtained using a Gaussian process regression introduced in Ref.~\cite{Harada11}. The regression assumes only the smoothness of the scaling function ${m^2(U,L) = L^{-2\beta/\nu} f_0^m[(U-U_c) L^{1/\nu}]}$, rather than a specific polynomial form and agrees within error bars with our estimate ${\beta/\nu = 0.766(6)}$ (cf. SM).

Following the same procedure we extract the anomalous dimension of the fermions $\eta_{\psi}$ from the off-diagonal elements of the single particle Green's function ${G_{ab}(\mathbf{k}) = \langle a_{\mathbf{k}}^{\dagger}\, b_{\mathbf{k}}^{\phantom{\dagger}}\rangle}$ with the FSS Ansatz ${G_{ab}(\mathbf{k}_\text{min})(R_{m^2},L) = L^{-\eta_{\psi}} f_0^G\!(R_{m^2})}$, where we again assume ${z=1}$ \cite{Campostrini14,Otsuka16}. At zero momentum, ${G_{ab}(\mathbf{0}) = 0}$, as ${G_{aa}(\mathbf{0}) = G_{bb}(\mathbf{0}) = 1/2}$ measures the local density per flavor. One has to resort to use the smallest lattice momentum ${\mathbf{k}_\text{min} = (2\pi/L,0)}$, where $\mathbf{k}_\text{min}\to 0$ in the TDL. Here the finite size corrections require us to neglect system sizes ${L<11}$ in order to avoid scaling corrections. The fit to the data in \cfig{fig:eta}(b) yields the estimate ${\eta_{\psi} = 0.177(1)}$.

\begin{figure}[t]
   \centering
   \includegraphics[width=\columnwidth]{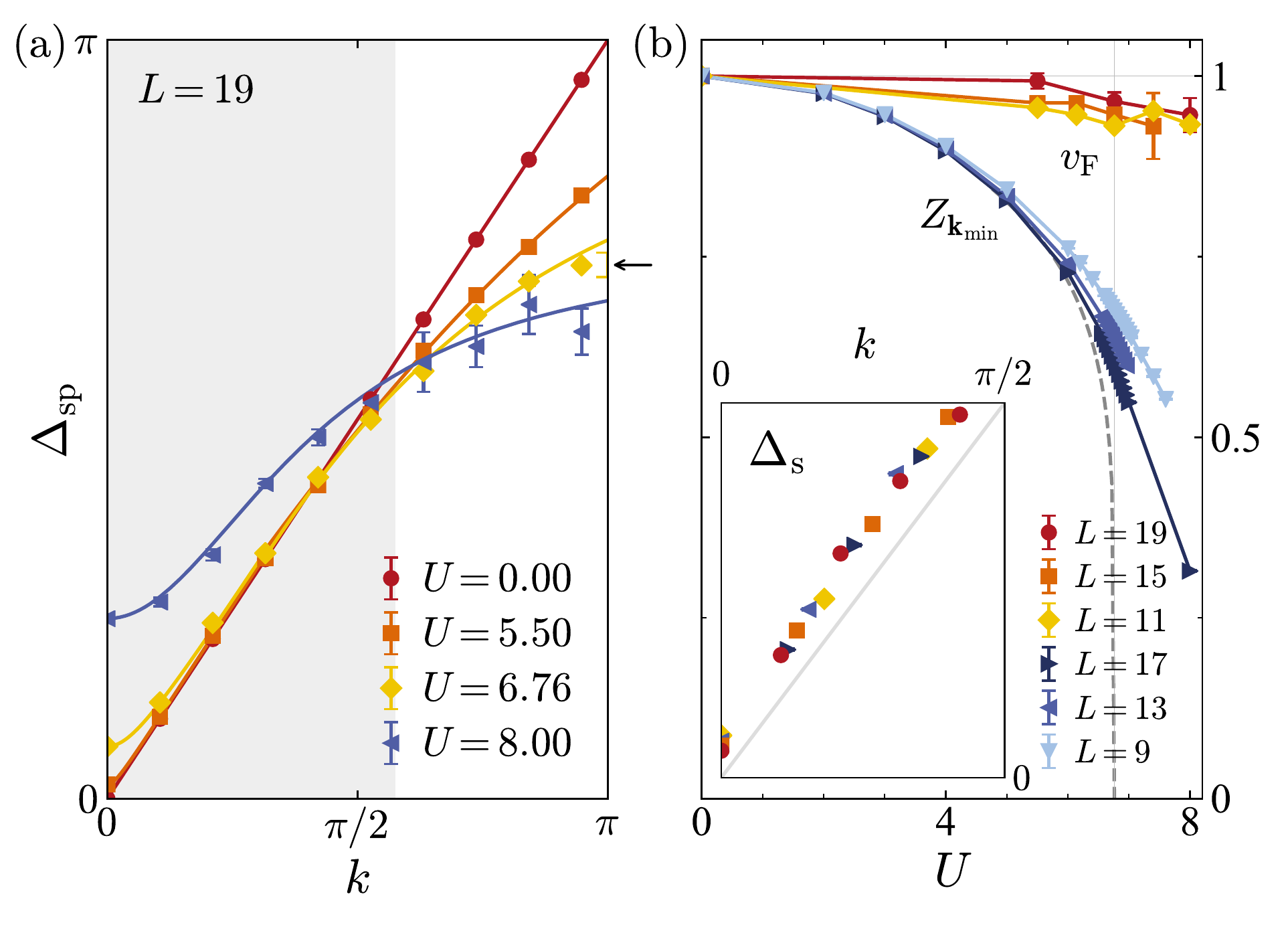}\vspace{-1em}
   \caption{(a) The single particle gap along $k_x$ ($k_y=0$) for different interactions strengths for an ${L=19}$ system. The shaded area corresponds to the data range from which the effective speed of light (equal to the Fermi velocity $v_\text{F}$ at $U=0$) has been extracted from a fit of the relativistic dispersion to the data points. Lines are guides to the eye only. (b) The quasi particle weight $Z_{\mathbf{k}_\text{min}}$ and the effective speed of light $v_\text{F}(U) $ for different system sizes. While increasing interaction strength degrades the quasi particle character, the Fermi velocity remains mostly unrenormalized up to the critical point. At large U we expect the speed of light to scale as $1/U$.}
   \label{fig:vF}
\end{figure}

The off-diagonal single particle Green's function is also proportional to the quasi particle weight (residue of the quasi particle pole) ${Z_{\mathbf{k}_\text{min}}=2G_{ab}(\mathbf{k}_\text{min})}$ \cite{Brinkman70,Fradkin13,Otsuka16}. As the critical point is approached from the noninteracting limit, growing correlations lead to increasing fluctuations in the semi-metal near the Fermi energy and the well defined fermionic quasi particle character of the chiral limit ${Z_{\mathbf{k}_\text{min}} = 1}$, is monotonously diminished ${Z_{\mathbf{k}_\text{min}}\to 0}$ as ${U\to U_c}$ \cite{Brinkman70,Herbut09a}. In order to show the consistency of our estimates, we plot the expected behavior of the residue of the quasi particle pole ${Z_{\mathbf{k}_\text{min}}\sim(U_c-U)^{\nu\eta_{\psi}}}$ in \cfig{fig:vF}(b) using the previously extracted exponents (dashed line). Beyond the critical point the Fermi point-surface is gapped out as ${\Delta_\text{sp}\sim(U-U_c)^{z\nu}}$ (not shown), and the fermionic primary excitations are replaced by the Goldstone bosons which originate from the spontaneous continuous symmetry breaking of the spin rotational symmetry in the TDL. 

\begin{figure*}[tp]
   \centering
	\includegraphics[width=0.333\textwidth]{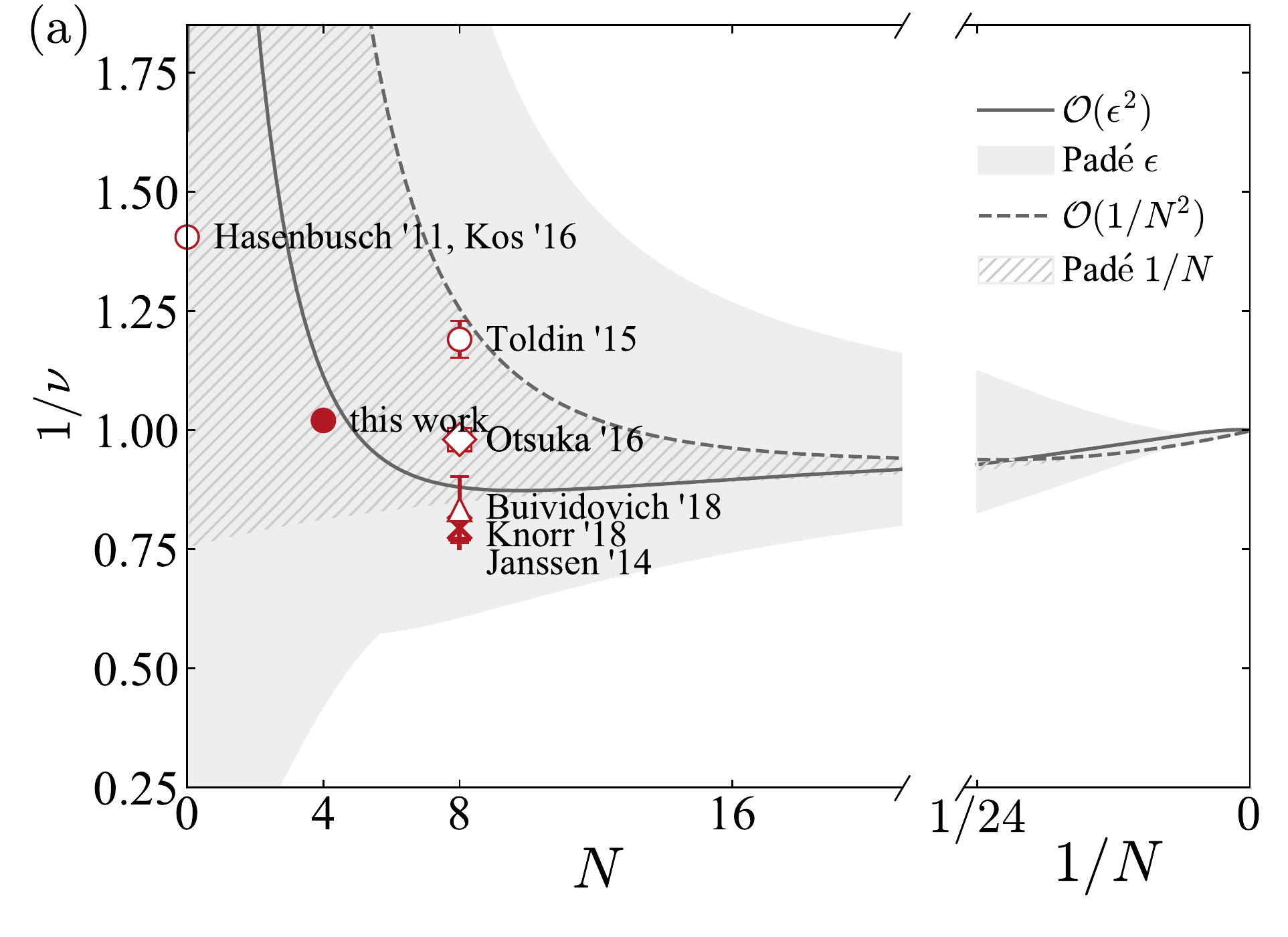}\hfill
	\includegraphics[width=0.333\textwidth]{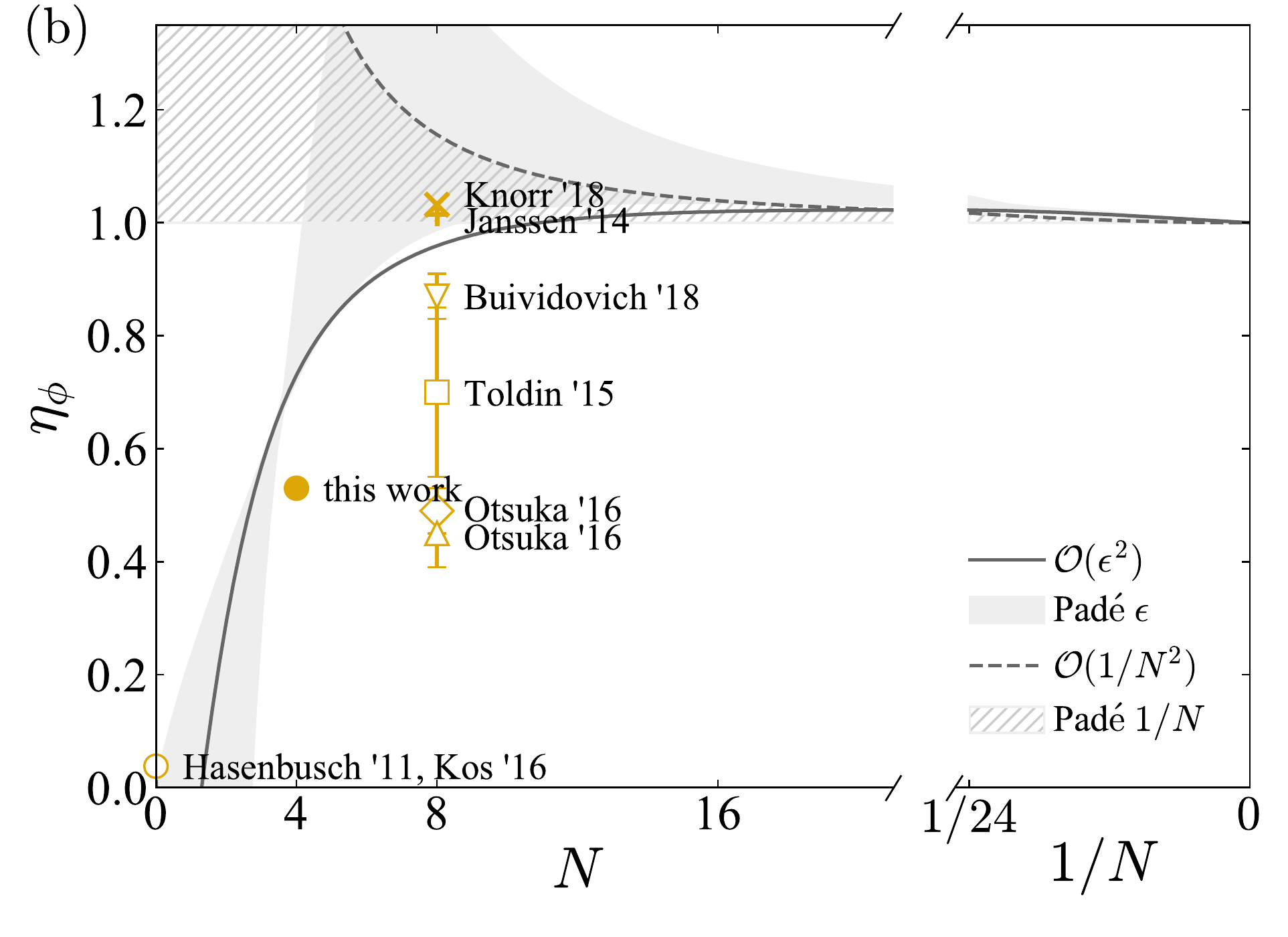}\hfill
	\includegraphics[width=0.333\textwidth]{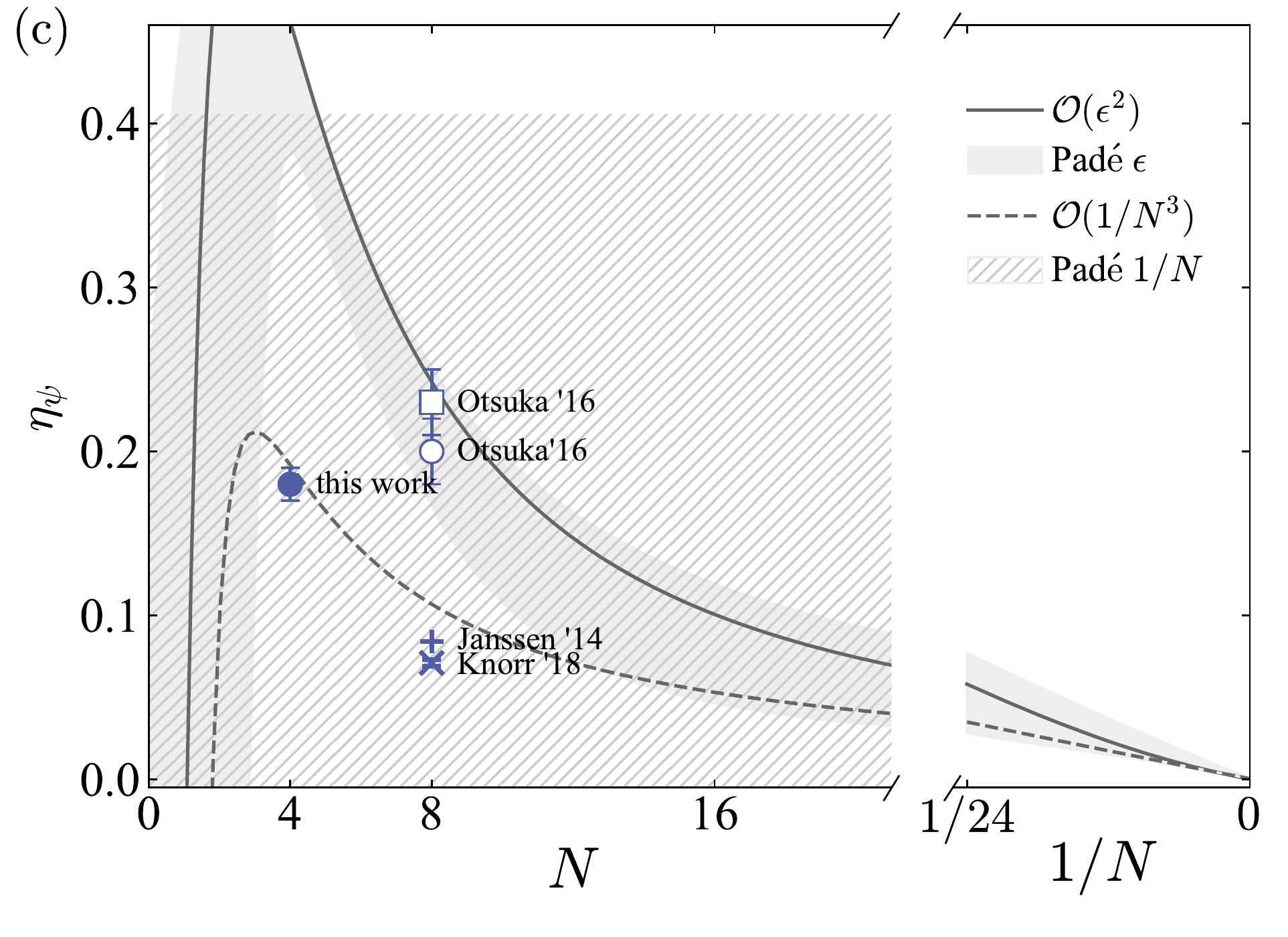}\vspace{-1em}\\
   \caption{Chiral Heisenberg GNY universality class: Comparison of estimates for (a) the correlation-length exponent $1/\nu$, (b) the boson anomalous dimension $\eta_{\phi}$ and (c) the fermion anomalous dimension $\eta_{\psi}$ for different numbers of fermion species, from Monte Carlo simulations (filled and open markers) \cite{Hasenbusch11,Toldin15,Otsuka18,Buividovich18}, conformal bootstrap \cite{Kos16}, functional RG ($+$, $\times$) \cite{Janssen14,Knorr18}, as well as fixed order expansions and Pad\'e approximants for the series from $\epsilon$-expansion \cite{Rosenstein93,Zerf17} and ${1/N}$-expansion \cite{Gracey18} (lines and shaded areas).
   \label{fig:comparison}}
\end{figure*}

In addition we determine the single particle gap $\Delta_\text{sp}(\mathbf{k})$ from a fit to the asymptotic long imaginary-time behavior of the single particle Green's function ${G_{aa}(\mathbf{k},\tau) = \langle a_{\mathbf{k}}^{\dagger}(\tau)\, a_{\mathbf{k}}^{\phantom{\dagger}}(0)\rangle \propto \exp[-\tau\Delta_\text{sp}(\mathbf{k})]}$ \cite{Feldbacher01}. In the SM we provide evidence for the relativistic finite size scaling ${\Delta_\text{sp} \sim L^{-z}}$ close to criticality, which validates our assumption that ${z=1}$. Cross sections of the momentum resolved excitation gap in \cfig{fig:sp_gap} are shown for different values of $U$ in \cfig{fig:vF}(a). The dynamically generated mass corresponds to ${\Delta_\text{sp}(k=0)}$ at vanishing momentum. The bandwidth decreases significantly with growing $U$, yet the single particle excitations close to the boundary of the Brillouin zone converge to a finite value at rather high energies, as indicated by the arrow for $U=6.76$ in the TDL. This implies that no additional zero modes are introduced by correlations \cite{Nason85,Gattringer10}. In order to study the impact of interactions on the low-energy dispersion, we fit the relativistic single particle dispersion ${\Delta_\text{sp}(k)=[\Delta_\text{sp}(0)^2 + (v_\text{F}(U) k)^2}]^{1/2}$ to the data for momenta within the grey shaded region to estimate the Fermi velocity $v_\text{F}$. This approach is validated by the expected spectrum both in the semimetallic and the symmetry broken phase. Exactly at the quantum critical point the spectrum is more complex~\cite{Schuler16,Sen15}. At the critical point the excitation velocity correspond to the speed of light of the conformal field theory. This procedure enables us to extract $v_\text{F}$ despite the vanishing quasi-particle weight at criticality. As illustrated in \cfig{fig:vF}(b) in the approach of the phase transition from the non-interacting limit the speed of light remains approximately constant. In our case ${v_\text{F}\approx v_\text{F}^0\approx 1}$, which is not necessarily the case in general -- the RG scaling Ansatz for the quasi-particle residue simply implies ${v_\text{F} \sim (U_c-U)^{\nu(z-1)}}$ to remain regular in the vicinity of the relativistic critical point \cite{Herbut09a}. Beyond the critical point the fermionic primary excitations are replaced by their bosonic counterpart such that close to $U_c$ the spin wave velocity ${v_{\phi} \approx v_\text{F}}$. The inset in \cfig{fig:vF}(b) illustrates the similar excitation velocities by comparing of the momentum dependence of the spin gap $\Delta_s(k) \propto v_{\phi}k$ with the Fermi velocity $v_{F}^0$ \cite{Roy16}.

\paragraph{Discussion ---}
Figure~\ref{fig:comparison} shows our results for the critical exponents in the context of recent results for different numbers of fermion species. For each case we have consistent estimates for the large-$N$ limit from $\epsilon$-expansion \cite{Rosenstein93,Zerf17} and $1/N$-corrections \cite{Gracey18}. The same holds for the limit $N=0$ where no fermions couple to the bosonic order parameter and high precision estimates from Monte Carlo simulations and conformal bootstrap calculations exist \cite{Hasenbusch11,Kos16}. For relatively small numbers of fermion species, ${N \lesssim 16}$, estimates from Monte Carlo (MC) simulations \cite{Toldin15,Otsuka18,Buividovich18} and analytic expansions differ significantly. For the analytical expansion results we include results at fixed expansion order ($\epsilon=1$) as indicated (lines), the range of values spanned by the Pad\'e approximant with all pole-free combinations of numerator and denominator order from one to the maximum expansion order available (shaded/hatched areas), as well as estimates from functional RG \cite{Janssen14,Knorr18}. The MC results generally follow the trends set by the analytical predictions, but for the anomalous dimension of the order parameter $\eta_{\phi}$ in the $1/N$ approximation. While the MC data are scattered, there appears to be a common trend for ${\nu,\eta_{\phi} \lesssim 1}$, which roughly follows the $\epsilon$-expansion at fixed ${\mathcal{O}(\epsilon^2)}$.

The non-monotonically decreasing coefficients of the series expansions \cite{Zerf17,Gracey18} are responsible for the large uncertainty at small $N$ and underscores the importance of future estimates and bounds from other approaches, such as conformal bootstrap calculations for the chiral Heisenberg transition. The tension between different results from MC simulations can be attributed to multiple sources: We can never exclude the possibility that the lattices sizes reached so far are simply not within the asymptotic scaling regime of the GNY transition, yet all MC investigations were able to extract critical exponents from a working FSS Ansatz. Also, different implementation of lattice fermions may avoid FSS contributions from non-leading irrelevant fields as has been seen recently in the context of quantum spin models \cite{Nvsen18}. Most importantly, the different maximum system sizes used, limit the momentum resolution of the relativistic dispersion at low energies. This is in contrast to SLAC fermions, which appears to be subject to smaller finite size corrections, which we further quantify for several correlation ratios in the SM.

\paragraph{Conclusion ---} 
We have presented the first QMC investigation of the critical properties of the $N=4$ chiral Heisenberg GNY quantum phase transition in 2+1$d$. To account for the ambiguity in the choice of the correlation ratio, fit ranges and included lattice sizes we report our conservative estimates for the critical exponents ${\nu = 0.98(1)}$, ${\eta_{\phi} = 0.53(1)}$, and ${\eta_{\psi} = 0.18(1)}$. The lattice realization of a single Dirac cone allowed us to significantly reduce finite size effects and access the regime of small fermion species numbers, which is essential to sort out the disparate results from complementary methods. Our approach opens the possibility to simulate the previously unexplored $N=2$ chiral Ising GNY transition and can be generalized to higher numbers of fermion flavors $N_f$. The single Dirac cone can be further generalized to anisotropic-, semi-, and birefringent Dirac semimetals \cite{Roy18a,Kennett11,Watanabe11,Roy18b}, which we leave to future investigations.

\begin{acknowledgments}
We thank S.~Hesselmann, L.~Janssen, C.~B. Lang, M.~Scherer, M.~Schuler and S.~Wessel for valuable discussions and comments. This research was supported by the Austrian Science Fund FWF the SFB FoQuS (F-4018). The computational results presented have been obtained using the HPC infrastructure LEO of the University of Innsbruck the Vienna Scientific Cluster VSC.
\end{acknowledgments}

%

\clearpage
\newpage

\onecolumngrid
\begin{center}\textbf{SUPPLEMENTAL MATERIAL}\end{center}
\vspace{\columnsep}
\twocolumngrid
\renewcommand\thefigure{S\arabic{figure}}    
\renewcommand\theequation{S\arabic{equation}}
\setcounter{figure}{0}    
\setcounter{equation}{0}
\setcounter{page}{1}
\setcounter{enumiv}{1}

\section*{A. Lattice fermions}

Let $N=2N_f$ denote the number of poles of the massless momentum space Dirac propagator $D$ in the continuum limit, each a fermionic one-particle state, or the number of zero modes on the lattice, respectively and $N_f$ be the number of generalized \textit{flavors} (also referred to as \textit{tastes} for staggered fermions) of each species of fermions. Then $N$ also corresponds to the number of components of the Dirac spinor representation, or the number of chiral Majorana modes. The chiral symmetry, besides particle conservation, as defined by the independent rotation of the chiral components ${D\gamma_5 = -\gamma_5D}$ then is ${{\rm SU}(N_f)_L \times {\rm SU}(N_f)_R}$. Where in the continuum field theory the special unitary groups are associated with left- and right-handed components of the spinor, in the Hubbard model they correspond to spin rotational and $\eta$-pairing symmetries of the Hamiltonian \cite{Fradkin13}. In contrast to spinless fermions ($N=2$) in 2+1$d$, the $N=4$ component spinors of electrons allow for a continuous rotation of the chiral components. Here, equivalent to the 3+1$d$ case, one can define a gamma matrix ${\gamma^5}$ that anti-commutes with all other $\gamma^{\mu}$ and the Hamiltonian in the massless limit, so that it becomes the generator for a continuous chiral symmetry \cite{Winkler15}.

\begin{figure}[!h]
    \centering
    \includegraphics[width=0.9\columnwidth]{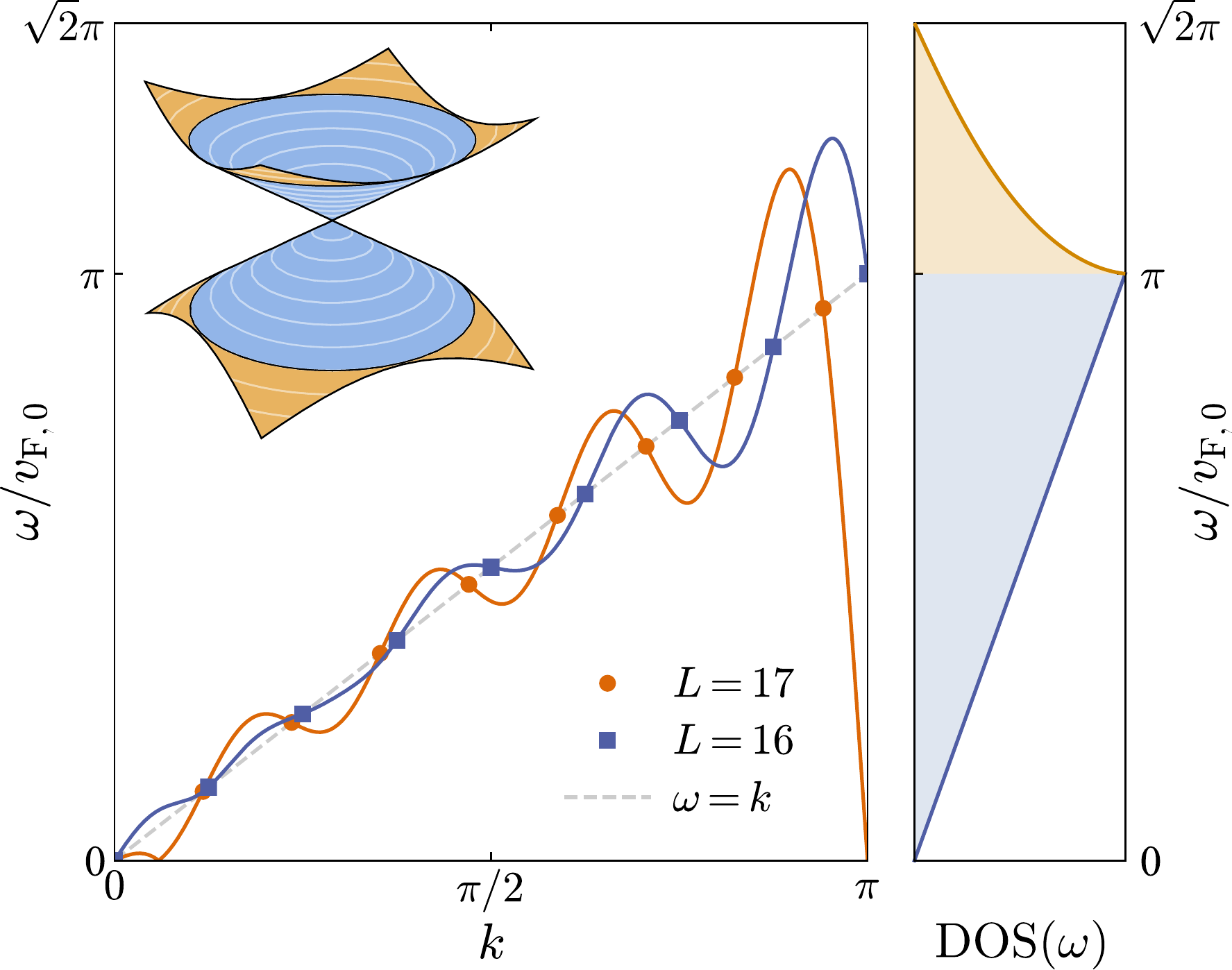}
    \caption{Although the continuous dispersion on finite size lattices exhibits severe ringing the linear energy-momentum relation is exact at the finite size lattice momenta as shown here for a system of odd and and even linear dimension $L$. In neither case exist doublers at the Brillouin zone boundary. The density of states (DOS) in the TDL is plotted alongside. The wingtips in the corners of the Brillouin zone, or rather the lack of centro-symmetric momenta, are responsible for the deviation from the linear DOS at high energies.}
    \label{fig:ekDOS}
\end{figure}

By construction, the relativistic dispersion is \textit{exactly} reproduced by \ceqn{Ht}, such that the eigenvalues of all finite size momenta obey ${\varepsilon(\mathbf{k}) = \pm v_\text{F}|\mathbf{k}|}$, independent of the system size. The eigenvalues at momenta not present on the finite size lattice however, show ringing which originates from the finite frequency Fourier transform and will strongly deviate from the linear form (cf. \cfig{fig:ekDOS}). At half-filling there exists a single two-fold degenerate point per flavor at zero frequency and there are no doublers at the Brillouin zone boundary. In contrast to Kogut-Susskind (staggered) fermions where the flavors are distributed among the sublattices, i.e., multiple Dirac cones exist at separate momenta, the SLAC fermions is the lattice realization closest to the continuum Dirac operator. 

%

\section*{B. Lorentz invariance}
\begin{figure}[!h]
  \centering
  \includegraphics[width=\columnwidth]{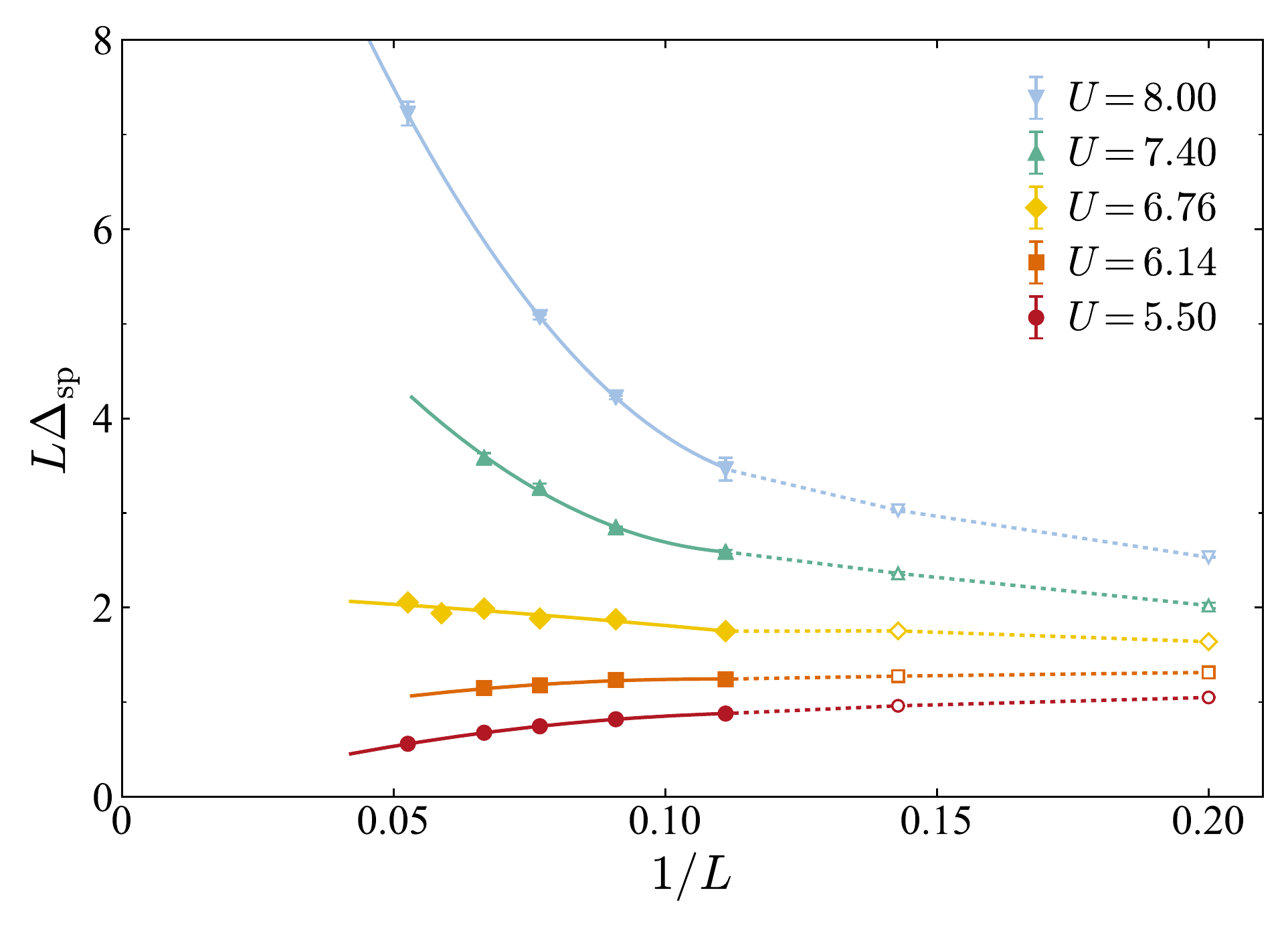}\vspace{-1em}
  \caption{The finite size extrapolation of $L$ times the single particle (mass) gap $\Delta_\text{sp}$ for different interaction strengths $U$. The extrapolation to a finite, non-zero value supports ${z=1}$ at the critical point ${U_c\approx 6.78}$. Data points with filled symbols have been fitted with a second order polynomial as a guide to the eye.}
  \label{fig:z}
\end{figure}

In order to justify the dynamical critical exponent ${z=1}$ at the critical point, we show the first fermionic excitation above the ground state (i.e., the mass gap), multiplied by the system size $L\Delta_\text{sp}$ in \cfig{fig:z}. The finite size extrapolation approaches a finite value and supports the relativistic finite size scaling ${\Delta_\text{sp} \sim L^{-z}}$ at criticality and diverges for ${U > U_c}$. Actually, by construction, from the noninteracting limit up to the critical point the semi-metal phase as described by the SLAC Hamiltonian implies a Lorentz invariant spectrum. The scaling of ${\Delta_\text{sp} \to 0}$ for ${U<U_c}$ is compatible with ${z=1}$ for simulations on a torus \cite{Schuler18}. At criticality the effective field theory which describes the system is the relativistic GNY field theory. Beyond the critical point, the fermions are gapped out and the low energy physics is captured by the O(3) nonlinear sigma model. Hence in the entire parameter regime $U\ge 0$ we never lose the relativistic property at low energies.
%

\section*{C. Finite size scaling analysis}

\subsection*{Correlation ratios}

\begin{figure}[t]
  \centering
  \includegraphics[width=\columnwidth]{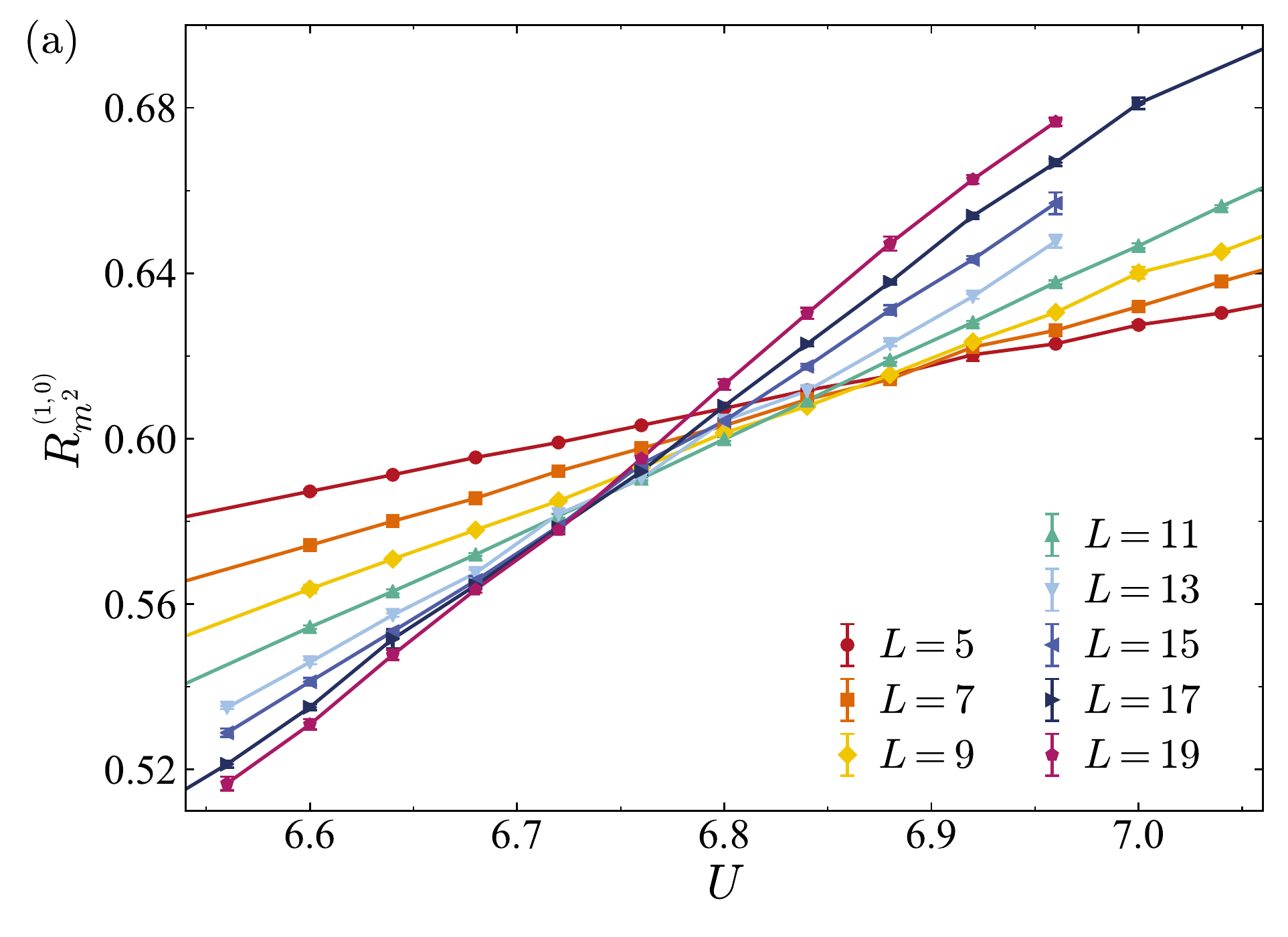}\\
  \includegraphics[width=\columnwidth]{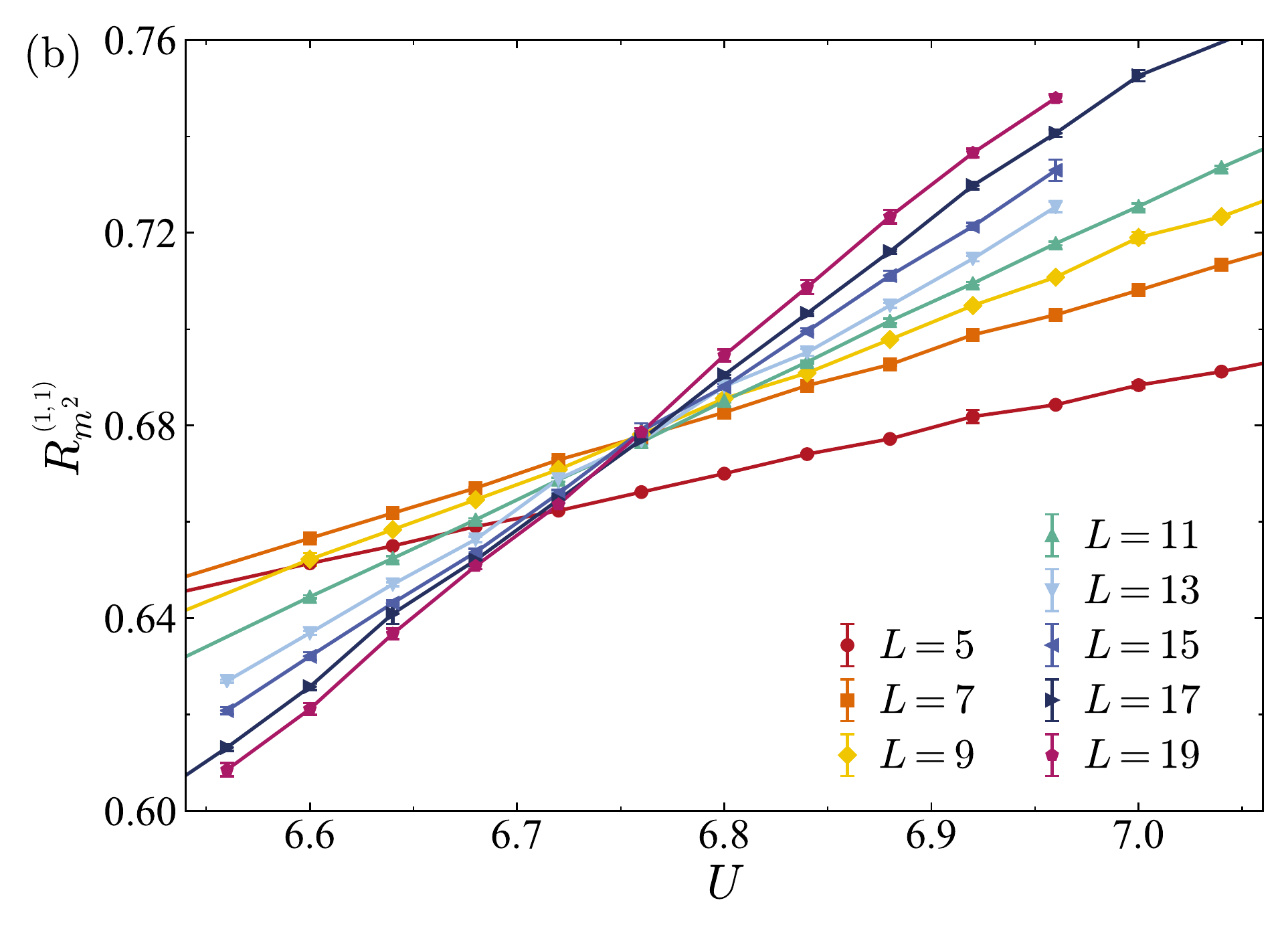}\\
  \includegraphics[width=\columnwidth]{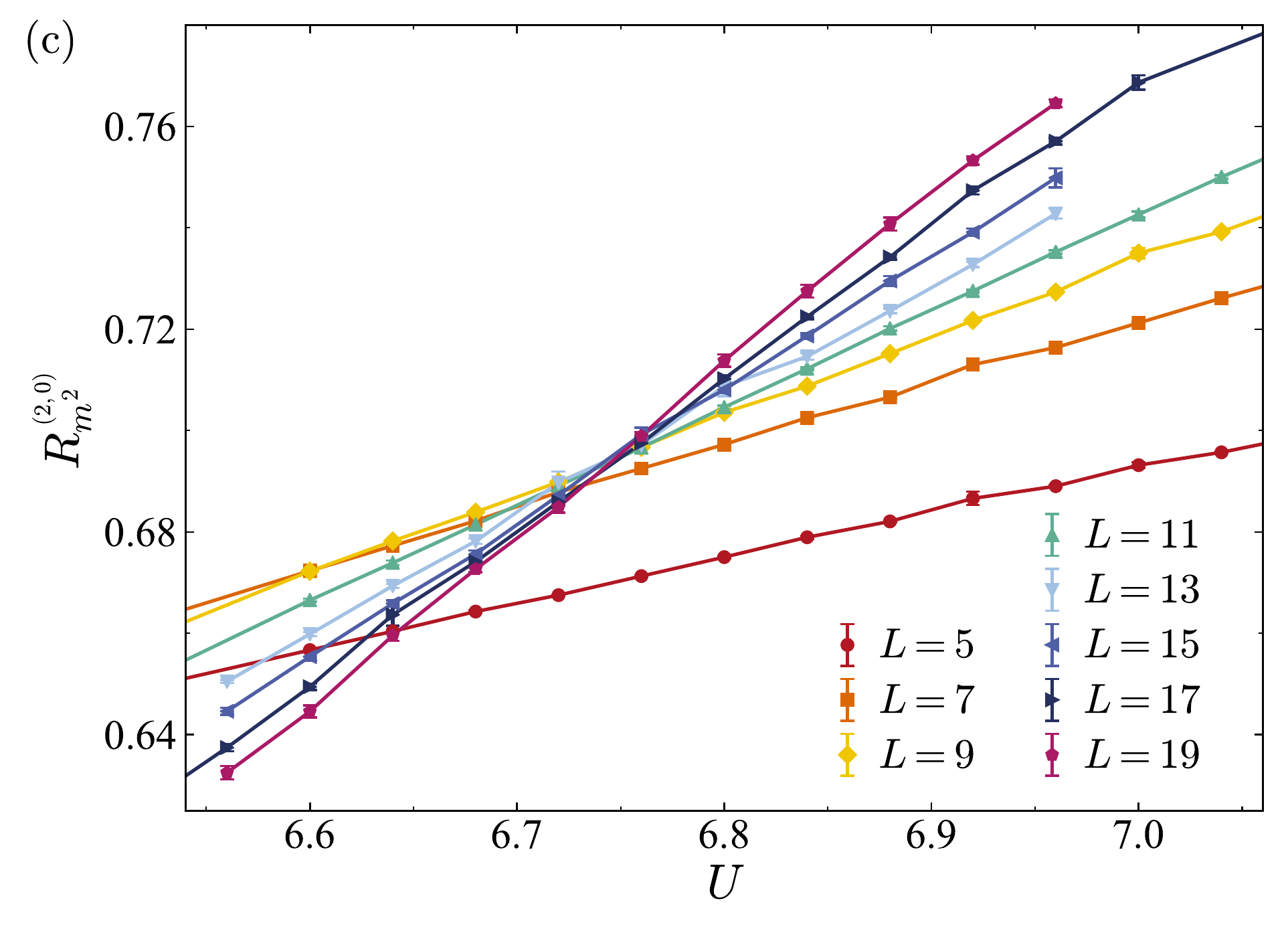}\vspace{-1em}
  \caption{The correlation ratios in the vicinity of the quantum phase transition. Lines are guides to they only. }
  \label{fig:corr_ratios}
\end{figure}

\begin{figure}[!tp]
  \centering
  \includegraphics[width=\columnwidth]{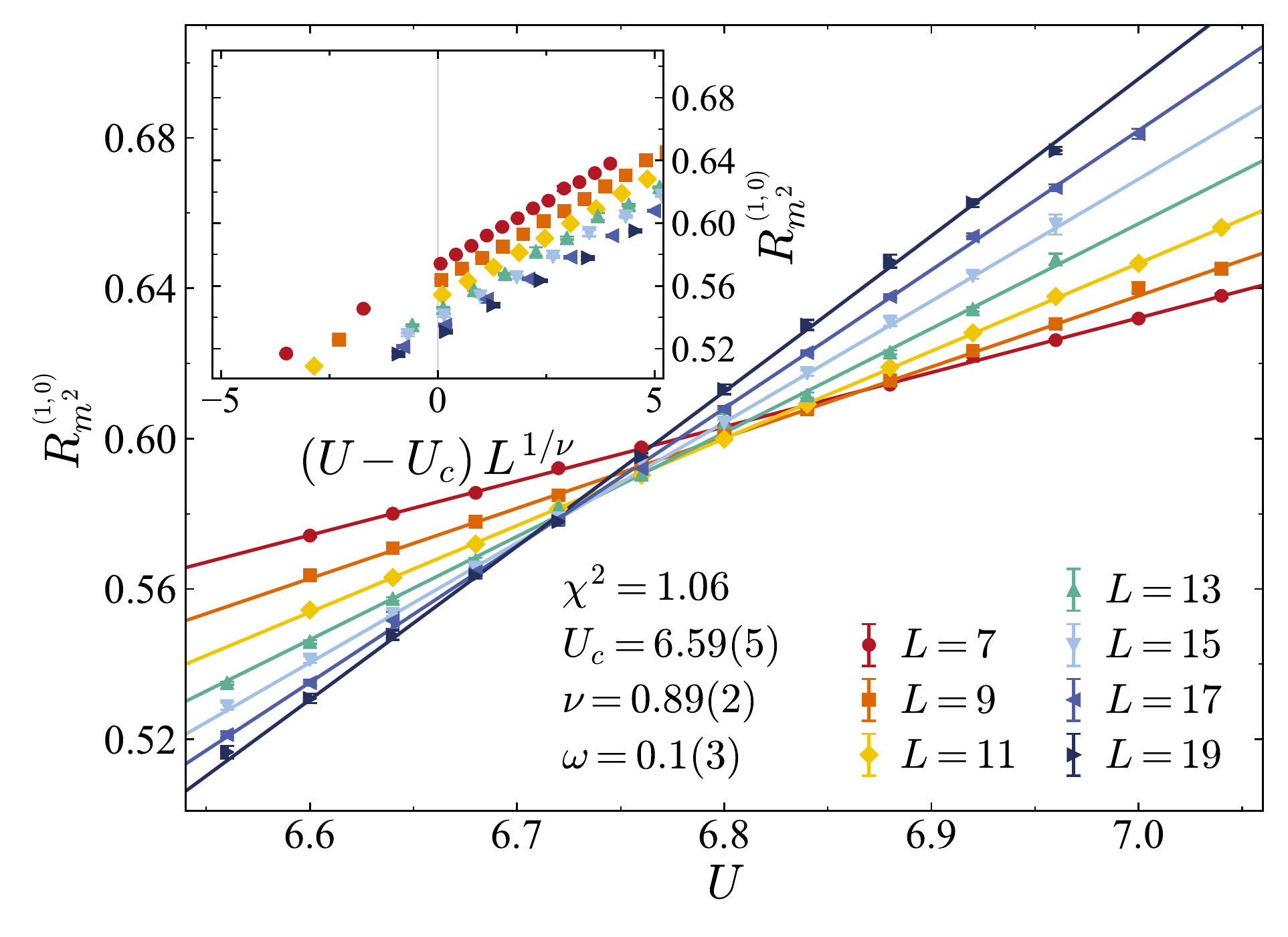}\vspace{-1em}
  \caption{The fit to the correlation ratio in \cfig{fig:corr_ratios}(a) and its data collapse (inset) in the vicinity of the quantum phase transition using the scaling Ansatz with corrections $(f_1\ne 0)$, and $n_\text{max}=2$. The breakdown of the data collapse invalidates the seemingly well fit.}
  \label{fig:corr_ratios_fail}
\end{figure}

The correlation ratios for different distances to the ordering momentum in $k$-space are shown in \cfig{fig:corr_ratios}(a)--(c). There exists a notable difference in the apparent drift in the various realizations. Fits to the ratio $R_{m^2}^{(1,0)}$ with the Ansatz
\begin{align}
    R(u,L) &= f_0^R\!\left(u  L^{1/\nu}\right) + L^{-\omega} f_1^R\!\left(u  L^{1/\nu}\right) \\
		     &= \sum_{n=0}^{n_\text{max}} a_n u^n\, L^{n/\nu} + L^{-\omega} \sum_{m=0}^{m_\text{max}} b_m u^m\, L^{m/\nu}\;,\nonumber
    \label{scalingAnsatz}
\end{align}
${u=U-U_c}$, do not yield satisfying results with ($f_1^R\ne 0$), or without ($f_1^R=0$) scaling corrections, independently of the included system sizes (cf. \cfig{fig:corr_ratios_fail}): Where the Ansatz allows for a decent fit to the data $R_{m^2}^{(1,0)}(U,L)$, the data do not collapse after rescaling of the axis ${U\to (U-U_c) L^{1/\nu}}$. Upon exclusion of the smallest system sizes, the scaling analysis of $R_{m^2}^{(2,0)}$ results become comparable to the more favorable ratio $R_{m^2}^{(1,1)}$, which has been used almost exclusively in the main text. The results for the correlation ratio $R_{m^2}^{(1,1)}$ without scaling corrections in the vicinity of the critical point are listed in \ctab{tab:Rm2_11_wo}. As more and more small systems are neglected, i.e., $L_\text{min} =5,7,\ldots$, the reduced $\chi^2$ approaches unity. The estimates are stable with respect to the expansion order, such that $n_\text{max}=2$ turns out to be sufficient, as higher order coefficients tend to vanish and no longer influence the fit. The inclusion of corrections to scaling yields comparable results at the cost of larger error bars (cf. \ctab{tab:Rm2_11_w}). The estimates for the exponent $\omega$ are highly variable and are either large, or the expansion coefficients $b_m$ vanish within statistical uncertainty -- both of which make the contribution of $f_1^R$ to scaling largely irrelevant. We thus conclude that the inclusion of corrections in the scaling Ansatz for $R_{m^2}^{(1,1)}$ is not appropriate.

In \ctab{tab:Rm2_11_Gauss_wo} and \ctab{tab:Rm2_11_Gauss_w} we present fit results of the Gaussian process regression of $R_{m^2}^{(1,1)}$ close to the critical point and over the wider range $U\in [2, 12]$. In contrast to the least squares fit of a polynomial of order $n_\text{max}=2$, this method is based on Bayesian statistics and assumes only the smoothness of a scaling function, i.e., it does not require a specific polynomial form \cite{Harada11}. While the regression without corrections to scaling produces comparable estimates in both ranges, correction to scaling do not improve the fits. Their inclusion results in an inconsistently small exponent $\omega$, which would suggest significant drift of the crossing points, which again is not supported by the quality of the corresponding data collapse.

\begin{table}
	\caption{Results of the fits with the scaling Ansatz $R_{m^2}^{(1,1)}(u,L) = f_0^R(u L^{1/\nu})$ for different expansion orders $n_\text{max}$ of $f_0$ in the range ${U\in [6.56, 7.04]}$.
	\label{tab:Rm2_11_wo}}
	\begin{tabular}{lr@{\hspace{1.2em}}r@{\hspace{1.5em}}l@{\hspace{1.4em}}l}\hline\hline\vspace{-1.45em}\\
		& $L_{\rm min}$ & \multicolumn{1}{l}{\hspace{0.5em}$\chi^2$} & \multicolumn{1}{l}{\hspace{0.75em}$U_c$} & \multicolumn{1}{l}{\hspace{0.75em}$\nu$} \\\hline
$n_\text{max}=1$ &  5 & 158.38 & 6.6956(6) & 1.022(4)  \\
                 &  7 & 3.01   & 6.7620(8) & 0.971(5)  \\
                 &  9 & 3.31   & 6.761(1)  & 0.969(8)  \\
                 & 11 & 1.81   & 6.751(2)  & 0.98(1)   \vspace{0.3em}\\
$n_\text{max}=2$ &  5 & 156.21 & 6.6902(7) & 0.997(4)  \\
                 &  7 & 2.69   & 6.7595(9) & 0.977(5)  \\
                 &  9 & 2.97   & 6.758(1)  & 0.982(8)  \\
                 & 11 & 1.60   & 6.749(2)  & 0.99(1)   \vspace{0.3em}\\
$n_\text{max}=3$ &  5 & 156.72 & 6.6936(7) & 0.962(5)  \\
                 &  7 & 2.71   & 6.7598(9) & 0.970(7)  \\
                 &  9 & 3.00   & 6.758(1)  & 0.975(10) \\
                 & 11 & 1.62   & 6.749(2)  & 0.99(1)   \\\hline\hline                 
	\end{tabular}
\end{table}

\begin{table}
	\caption{Results of the fits with the scaling Ansatz $R_{m^2}^{(1,1)}(u,L) = f_0^R(u L^{1/\nu}) + L^{-\omega}\,f_1^R(u L^{1/\nu})$ for expansion order ${n_\text{max} = 2}$ of $f_0^R$ and different expansion orders $m_\text{max}$ of $f_1^R$ in the range ${U\in [6.56, 7.04]}$.}
	\begin{tabular}{lr@{\hspace{1.2em}}r@{\hspace{1.5em}}l@{\hspace{1.4em}}l@{\hspace{0.7em}}l}\hline\hline\vspace{-1.45em}\\
		& $L_{\rm min}$ & \multicolumn{1}{l}{\hspace{0.5em}$\chi^2$} & \multicolumn{1}{l}{\hspace{0.75em}$U_c$} & \multicolumn{1}{l}{\hspace{0.75em}$\nu$} & \multicolumn{1}{c}{\hspace{0.em}$\omega$} \vspace{0.1em}\\\hline
$m_\text{max}=0$ & 5  & 2.56 & 6.759(1) & 0.977(4)  & 33(2103) \\
                 & 7  & 2.24 & 6.73(6)  & 0.974(10) & 0(2)     \\
                 & 9  & 1.09 & 6.7(2)   & 0.96(2)   & 0(2)     \\
                 & 11 & 0.71 & 6.730(9) & 0.98(1)   & 9(8)     \vspace{0.3em}\\
$m_\text{max}=1$ & 5  & 2.60 & 6.759(1) & 0.977(5) & 19(18) \\
                 & 7  & 2.18 & 6.73(2)  & 1.3(4)   & 0.0(5) \\
                 & 9  & 1.00 & 6.6(2)   & 2(2)     & 0(1)   \\
                 & 11 & 0.71 & 6.73(1)  & 1.01(4)  & 8(7)   \vspace{0.3em}\\
$m_\text{max}=2$ & 5  & 2.56 & 6.760(1) & 0.977(5) & 32(1741) \\
                 & 7  & 2.11 & 6.74(4)  & 0.7(3)   & 0(1) \\
                 & 9  & 1.03 & 6.6(3)   & 0.7(5)   & 0(2) \\
                 & 11 & 0.71 & 6.73(1)  & 1.01(4)  & 8(6)   \\\hline\hline
	\end{tabular}
	\label{tab:Rm2_11_w}
\end{table}

\begin{table}
	\caption{Results of the Gaussian process regression with and without scaling corrections for $R_{m^2}^{(1,1)}$ in the range ${U\in [6.56, 7.04]}$.}
	\begin{tabular}{lr@{\hspace{1.2em}}r@{\hspace{1.5em}}l@{\hspace{1.4em}}l@{\hspace{0.7em}}l}\hline\hline\vspace{-1.45em}\\
		& $L_{\rm min}$ & \multicolumn{1}{l}{\hspace{0.5em}$\chi^2$} & \multicolumn{1}{l}{\hspace{0.75em}$U_c$} & \multicolumn{1}{l}{\hspace{0.75em}$\nu$} & \multicolumn{1}{c}{\hspace{0.em}$\omega$} \vspace{0.1em}\\\hline
w/o corr. & 5  & 220.35 & 6.730(5)  & 0.99(4) & $               $ \\ 
          & 7  &   2.97 & 6.758(2)  & 0.97(1) & $               $ \\
          & 9  &   3.56 & 6.756(2)  & 0.96(2) & $               $ \\
          & 11 &   2.04 & 6.747(2)  & 0.98(2) & $               $ \vspace{0.3em}\\
w/  corr. & 5  &  61.41 & 6.85(1)   & 0.91(2) & $     -0.083(10)$ \\ 
          & 7  &   2.78 & 6.714(10) & 0.99(1) & $\pmin 0.030(6) $ \\ 
          & 9  &   2.05 & 6.64(1)   & 1.03(2) & $\pmin 0.086(9) $ \\ 
          & 11 &   0.86 & 6.57(3)   & 1.09(2) & $\pmin 0.13(2)  $ \\\hline\hline
	\end{tabular}
	\label{tab:Rm2_11_Gauss_wo}
\end{table}

\begin{table}[h]
	\caption{Results of the Gaussian process regression for $R_{m^2}^{(1,1)}$ in the range ${U\in [2, 12]}$.}
	\begin{tabular}{r@{\hspace{1.2em}}r@{\hspace{1.5em}}l@{\hspace{1.4em}}l}\hline\hline\vspace{-1.45em}\\
		$L_{\rm min}$ & \multicolumn{1}{l}{\hspace{0.5em}$\chi^2$} & \multicolumn{1}{l}{\hspace{0.75em}$U_c$} & \multicolumn{1}{l}{\hspace{0.75em}$\nu$}  \vspace{0.1em}\\\hline
      5  & 5693.70 & 6.79(1)  & 0.661(8) \\                      
      7  &   20.67 & 6.765(3) & 0.868(4) \\                      
      9  &    8.42 & 6.764(3) & 0.867(6) \\                      
      11 &    4.79 & 6.756(3) & 0.861(6) \\\hline\hline
	\end{tabular}
	\label{tab:Rm2_11_Gauss_w}
\end{table}

\begin{figure}[tp]
  \centering
  \includegraphics[width=\columnwidth]{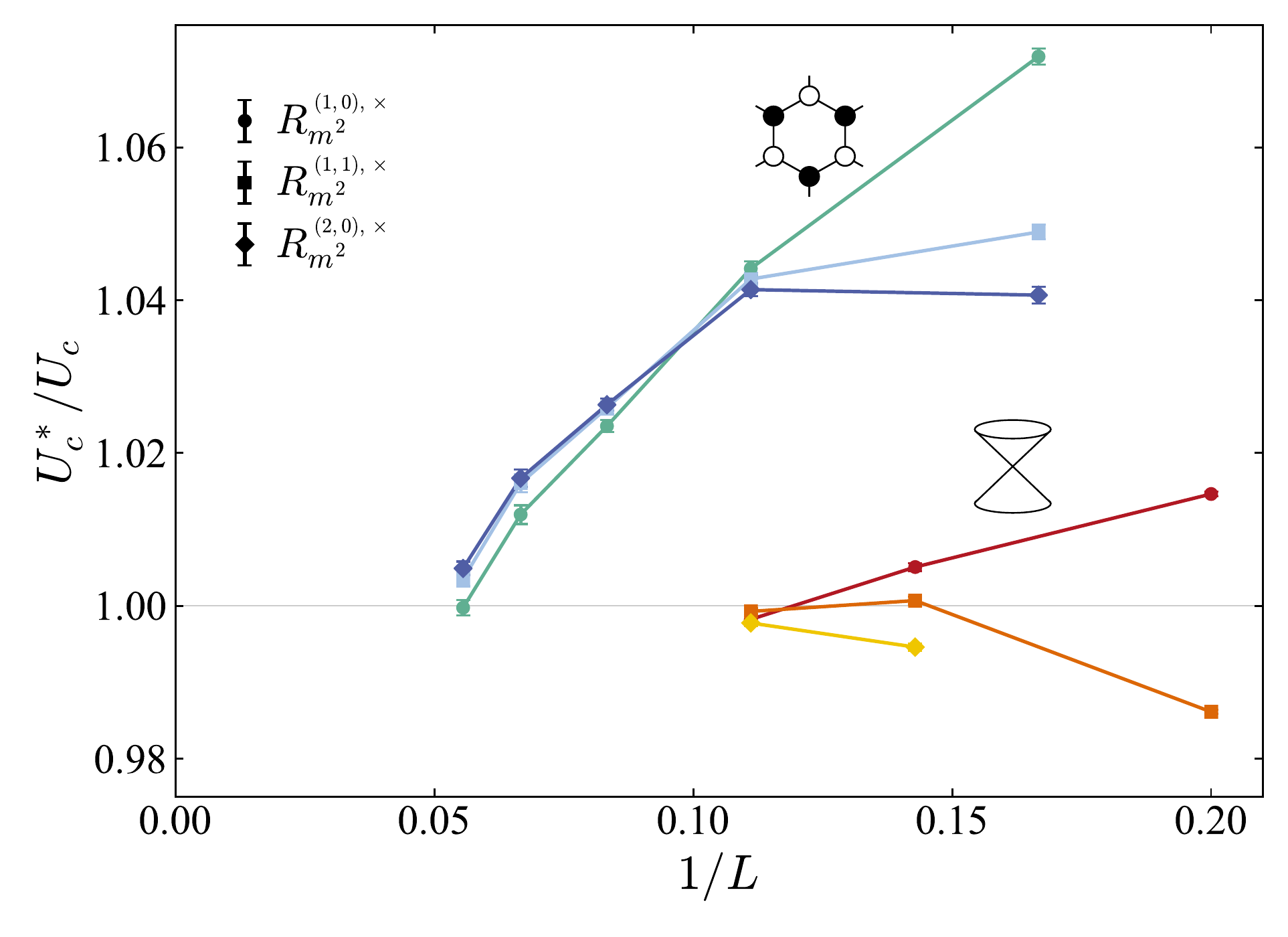}\vspace{-1em}
  \caption{The crossing points from the correlation ratios ${R_{m^2}^{(n_1,n_2)}}$ of $(L,2L+1)$ and $(L,2L)$ systems for the Dirac and honeycomb lattice. To allow the comparison of their distance with respect to their critical points, both have been normalized by $U_c=6.76$, $U_c=3.78$ \cite{Otsuka16}, respectively.}
  \label{fig:crossing}
\end{figure}

In \cfig{fig:crossing} we compare the crossing points of the correlation ratios $R_{m^2}^{(1,0)}$, $R_{m^2}^{(1,1)}$ and $R_{m^2}^{(2,0)}$ for SLAC fermions and the honeycomb lattice as they approach the critical coupling. We choose the crossings of finite size lattices $(L,2L+1)$ up to ${L=19}$ and $(L,2L)$ up to ${L=36}$, respectively. The vertical distance to the critical point, which has been normalized to one using $U_c=6.76$ and $U_c=3.78$ respectively, indicates the finite size dependence and the need for finite size scaling corrections. The SLAC fermions appear to be significantly less affected by finite size lattice effects, which suggests that significantly smaller lattices are needed to achieve a comparable accuracy in the determination of the critical exponents.

\subsection*{Magnetization}
\begin{figure}[t]
   \centering
   \includegraphics[width=\columnwidth]{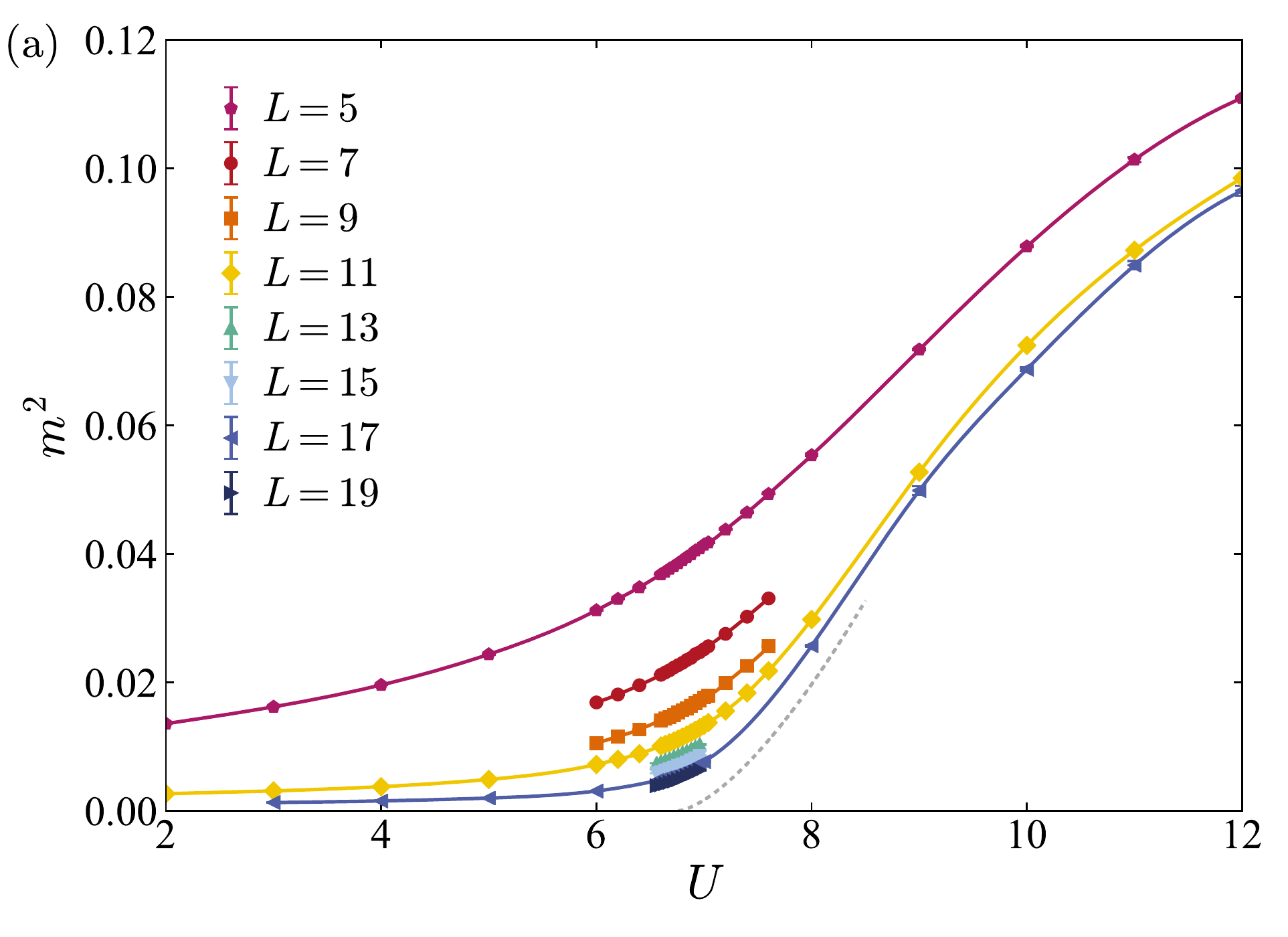}\\
   \includegraphics[width=\columnwidth]{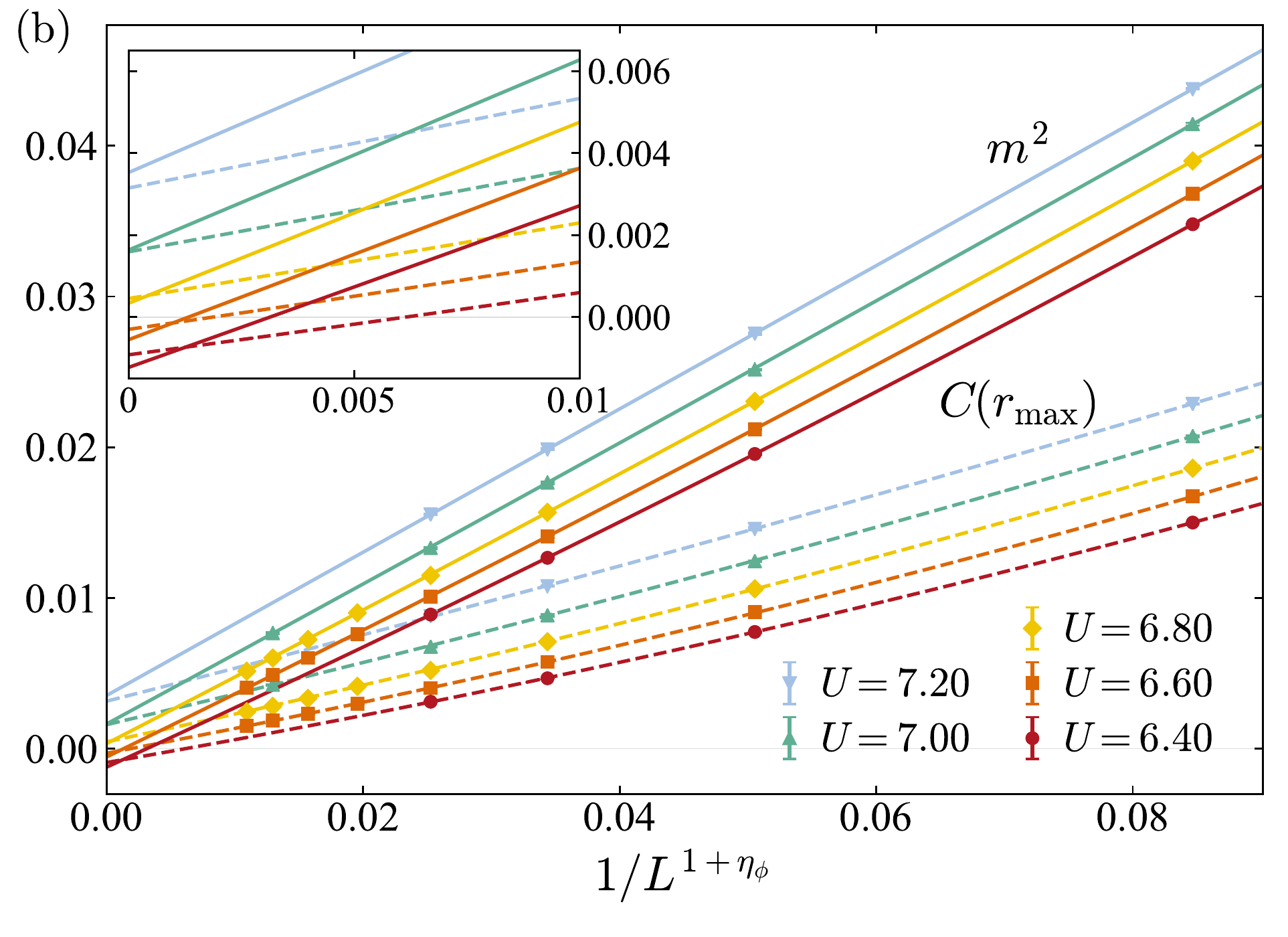}\vspace{-1em}
   \caption{(a) Squared magnetization $m^2$ versus coupling strength $U$ for different system sizes. The dotted line corresponds to onset of $m^2$  using the critical exponent $\beta$ as obtained from $\eta$ and $\nu$ via the hyperscaling relations. (b) The finite size extrapolation of $m^2$ and the spin correlation function at the largest finite size distance close to the phase transition.}
   \label{fig:m2}
\end{figure}

A scan of the squared magnetization $m^2$ as a function of the coupling strength $U$ for different system sizes is presented in \cfig{fig:m2}(a) alongside the finite size extrapolation of $m^2$ and the spin correlation function at the largest distance close to the phase transition in \cfig{fig:m2}(b). The dotted line in (a) corresponds to onset of ${m^2 \sim (U-U_c)^{2\beta}}$ in the TDL, where we use the critical coupling $U_c$ and exponent ${\beta=\frac{\nu}{2}(1+\eta_{\phi})}$, as obtained via the hyperscaling relation and the critical exponents extracted from the data in Figures~\ref{fig:nu} and \ref{fig:eta}. The fit results for $\eta_{\phi}$ are listed in \ctab{tab:eta_phi}. The finite size extrapolations in \cfig{fig:m2}(b) shows that the $m^2$ vanishes faster than ${L^{-(1+\eta)}}$ below ${U\le 6.60}$ and extrapolates to a finite value for ${U \ge 6.80}$, which is in accordance with our estimate of ${U_c \approx 6.76}$. Rescaling the axes of \cfig{fig:m2}(a) according to the Ansatz ${m^2(U,L) = L^{-(1+\eta_{\phi})}\, f_0^m[(U-U_c) L^{1/\nu}]}$, leads to the finite size scaling collapse shown in \cfig{fig:nu}(b) and validates the previously extracted critical exponents. A polynomial fit to the data provides an estimate for the scaling function (dotted line).

\begin{table}[h]
	\caption{Results of the fits with the scaling Ansatz $m^2(R,L) = L^{-(1+\eta_{\phi})} f_0^m(R_{m^2}^{(1,1)},L)$ for different orders $n_\text{max}$ in the range ${U\in [6.56, 7.04]}$.}
	\begin{tabular}{lr@{\hspace{1.2em}}r@{\hspace{1.5em}}l}\hline\hline\vspace{-1.45em}\\
		& $L_{\rm min}$ & \multicolumn{1}{l}{\hspace{0.5em}$\chi^2$} & \multicolumn{1}{l}{\hspace{0.75em}$\eta_{\phi}$} \vspace{0.1em}\\\hline
$n_\text{max}=1$ &  5 & 179.40 &  0.5928(3) \\
                 &  7 &  30.71 &  0.5326(6) \\
                 &  9 &  27.62 &  0.514(1)  \\
                 & 11 &  28.07 &  0.511(2)  \vspace{0.3em}\\
$n_\text{max}=2$ &  5 & 131.10 &  0.6071(4) \\
                 &  7 &   5.53 &  0.5469(7) \\
                 &  9 &   0.97 &  0.531(1)  \\
                 & 11 &   0.83 &  0.530(2)  \vspace{0.3em}\\
$n_\text{max}=3$ &  5 & 128.00 &  0.6059(4) \\
                 &  7 &   4.63 &  0.5472(7) \\
                 &  9 &   0.41 &  0.532(1)  \\
                 & 11 &   0.28 &  0.530(2)  \\\hline\hline
	\end{tabular}
	\label{tab:eta_phi}
\end{table}

\begin{table}[h]
	\caption{Results of the Gaussian process regression for the scaling Ansatz ${m^2(u,L) = L^{-2\beta/\nu} f_0^m(u L^{1/\nu})}$ in the range ${U\in [6.60, 6.96]}$.}
	\begin{tabular}{r@{\hspace{1.2em}}r@{\hspace{1.5em}}l@{\hspace{1.4em}}l@{\hspace{0.7em}}l}\hline\hline\vspace{-1.45em}\\
		$L_{\rm min}$ & \multicolumn{1}{l}{\hspace{0.5em}$\chi^2$} & \multicolumn{1}{l}{\hspace{0.75em}$U_c$} & \multicolumn{1}{l}{\hspace{0.75em}$\nu$} & \multicolumn{1}{c}{\hspace{0.em}$\beta/\nu$} \vspace{0.1em}\\\hline
 5 & 1.69 & 6.660(4) & 0.949(9) & 0.766(2) \\
 7 & 1.26 & 6.648(7) & 0.98(1)  & 0.753(4) \\
 9 & 1.24 & 6.64(1)  & 0.97(1)  & 0.73(1)  \\
11 & 0.95 & 6.60(3)  & 1.01(2)  & 0.68(2)  \\\hline\hline
	\end{tabular}
	\label{tab:m2_Gauss}
\end{table}

In \ctab{tab:m2_Gauss} we present fit results of the Gaussian process regression close to the critical point. In contrast to the regression of the correlation ratio, which coincides with the polynomial fits, here the critical coupling deviates. While this affects the spread of the data along the $u$-axis the shape (cf. \cfig{fig:nu}) is determined by $\beta/\nu$, which agrees well with our previous estimate ${\beta/\nu = 0.766(6)}$.

\subsection*{Anomalous dimension of the fermions}
The finite size scaling analysis for the fermion anomalous dimensions $\eta_{\psi}$ as described in the main text, for different polynomial expansion orders are presented in \ctab{tab:eta_psi}. The expansion order $n_\text{max}=2$ yields stable results upon exclusion of the smallest system sizes.

\begin{table}[t]
	\caption{Results of the fits of $G_{ab} = L^{-\eta_{\psi}} f_0^G(R_{m^2}^{(1,1)},L)$ for different orders $n_\text{max}$ in the range $U\in [6.56, 7.04]$.}
	\begin{tabular}{lr@{\hspace{1.2em}}r@{\hspace{1.5em}}l}\hline\hline\vspace{-1.45em}\\
		& $L_{\rm min}$ & \multicolumn{1}{l}{\hspace{0.5em}$\chi^2$} & \multicolumn{1}{l}{\hspace{0.75em}$\eta_{\psi}$} \vspace{0.1em}\\\hline
$n_\text{max}=1$ &  7 & 27.10 & 0.1665(2)   \\
                 &  9 &  6.77 & 0.1777(4)   \\
                 & 11 &  6.58 & 0.1810(6)   \\
                 & 13 &  7.34 & 0.1826(9)   \vspace{0.3em}\\
$n_\text{max}=2$ &  7 & 18.87 & 0.1630(2)   \\
                 &  9 &  2.28 & 0.1746(4)   \\
                 & 11 &  1.72 & 0.1774(7)   \\
                 & 13 &  1.96 & 0.1789(9)   \vspace{0.3em}\\
$n_\text{max}=3$ &  7 & 18.26 & 0.1632(2)   \\
                 &  9 &  2.32 & 0.1746(4)   \\
                 & 11 &  1.73 & 0.1774(7)   \\
                 & 13 &  1.91 & 0.1791(10)  \\\hline\hline
	\end{tabular}
	\label{tab:eta_psi}
\end{table}

\subsection*{Summary}
For convenience we summarize the critical exponents computed in this manuscript along with their associated scaling dimensions:
\begin{equation*}
   \begin{split}
      1/\nu &= 1.02(1)\;,       \phantom{\frac{1}{2}}\\
      \eta_{\phi} &= 0.53(1)\;, \phantom{\frac{1}{2}}\\
      \eta_{\psi} &= 0.18(1)\;, \phantom{\frac{1}{2}}
   \end{split}\quad\quad
   \begin{split}
      \Delta_{\epsilon} &= 3-1/\nu = 1.98(1)\;,  \phantom{\frac{1}{2}}\\
      \Delta_{\phi} &= \frac{1}{2}(\eta_{\phi}+1) = 0.765(5)\;, \\
      \Delta_{\psi} &= \frac{1}{2}(\eta_{\psi}+2) = 1.090(5)\;.
   \end{split}
\end{equation*}

\section*{D. Simulation details}
\begin{figure}[b]
   \centering
   \includegraphics[width=\columnwidth]{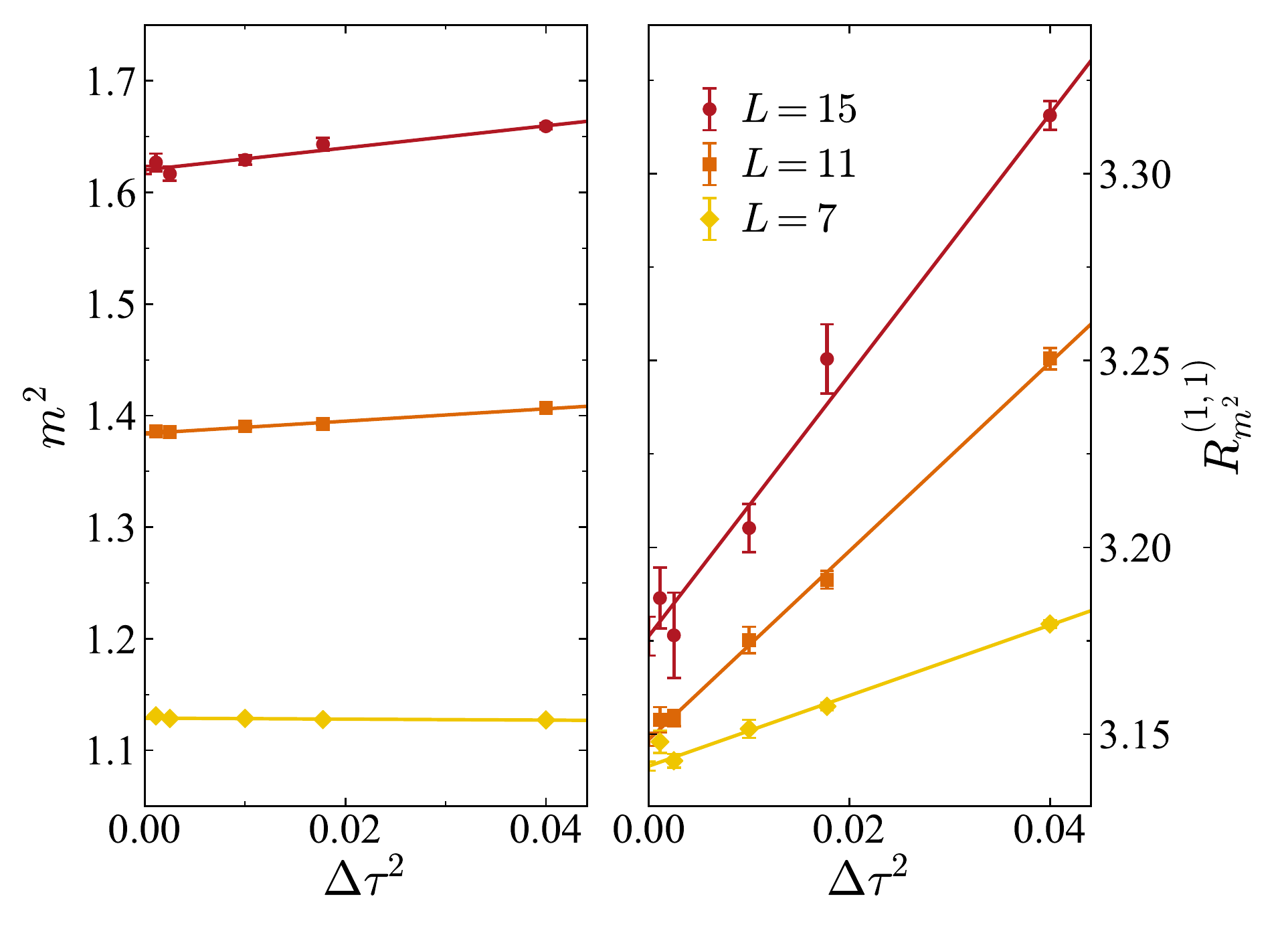}\vspace{-1em}
   \caption{The finite size extrapolation of the squared magnetization and the correlation ratio into the continuous imaginary time limit close to criticality at ${U=6.76}$.}
   \label{fig:dtau}
\end{figure}

All results were obtained from projective (${T=0}$) auxiliary-field (determinantal) QMC simulations based on an SU(2) symmetric Hubbard-Stratonovich decomposition, such that the auxiliary spins couple to the charge density \cite{Sugiyama86,Assaad08}. Observables were measured according to 
\begin{align}
	\langle\mathcal{O}\rangle = \frac{\langle\Psi_\text{T}|\E^{-\theta H}\mathcal{O}\E^{-\theta H}|\Psi_\text{T}\rangle}{\langle\Psi_\text{T}|\E^{-2\theta H}|\Psi_\text{T}\rangle}\;,
\end{align}
using the equal-time and imaginary time-displaced single-particle Green function and Wick's theorem \cite{Feldbacher01}, the free (massless) system acts as the trial wave function $|\Psi_\text{T}\rangle$ and $\theta$ denotes the projection length. Imaginary time was discretized with a Trotter time step ${\Delta\tau = \beta/N_{\tau} \le 0.1}$, where $N_{\tau}$ denotes the number of time slices. We chose projections of $2\theta=40$ and 70 for simulations of equal-time and time-displaced measurements, respectively, and checked the convergence of our results within their statistical uncertainty. A symmetric Suzuki Trotter decomposition 
\begin{align}
   \E^{-2\Theta H} = \left[ \E^{-\frac{1}{2}\Delta\tau H_t} \E^{-\Delta\tau H_U} \E^{-\frac{1}{2}\Delta\tau H_t} \right]^{N_{\tau}}\;,
\end{align}
was employed, which results in an error of ${\mathcal{O}(\Delta\tau^3)}$ in the short time propagation, or ${\mathcal{O}(\Delta\tau^2)}$ for observables. The impact of the Trotter error is illustrated for the squared magnetization and the correlation ratio close to criticality at ${U=6.76}$ in \cfig{fig:dtau}. The finite size results for the correlation ratio monotonously decrease as a function of $\Delta\tau$ and monotonously increase as a function $L$, indicating the absence of a crossover scale for the observed system sizes. The discrete imaginary time Trotter error acts an ultraviolet regularization, hence contributes to the physics at high energies and is as such not expected to affect the critical exponents significantly.

In \ctab{tab:ED} we show QMC estimates of the total energy ${\langle H\rangle}$ and the double occupancy ${\langle n_d\rangle = \sum_{i,c\in\{a,b\}} \langle n_{ic1} n_{ic2}\rangle}$, where ${n_{ic\sigma} = c_{i\sigma}^{\dagger} c_{i\sigma}^{\phantom{\dagger}}}$ for an 18-site ($L=3$) system at ${U=3}$, 6 and different values of imaginary time discretization $\Delta\tau$. The extrapolation of the data to the continuous imaginary time limit ${\Delta\tau\to 0}$ coincide with the exact diagonalization results within error bars.

\begin{table}
   \caption{Comparison of QMC estimates of the total energy and the double occupancy with exact diagonalization results for different values of imaginary time discretization and its extrapolation ${\Delta\tau\to 0}$ on a $L=3$ (18-sites) cluster at $U=3$ and $U=6$.}
	\begin{tabular}{c@{\hspace{1.5em}}c@{\hspace{1.5em}}l@{\hspace{1.5em}}l}\hline\hline\vspace{-1.5em}\\
      & \multicolumn{1}{c}{$\Delta\tau$}\hspace{0.5em} & \multicolumn{1}{c}{$\!\!\!\!\!\!-\langle H\rangle$}\hspace{1.4em} & \multicolumn{1}{c}{$\langle n_d\rangle$} \\\hline
      $U = 3$ & 0.4000  & 9.839(3)  & 2.6729(9) \\
              & 0.3333  & 9.880(4)  & 2.597(1)  \\
              & 0.2667  & 9.902(3)  & 2.5333(9) \\
              & 0.2000  & 9.914(2)  & 2.4821(6) \\
              & 0.1333  & 9.918(1)  & 2.4445(5) \\
              & 0.1000  & 9.917(2)  & 2.4318(5) \\
              & 0.0500  & 9.920(3)  & 2.4183(8) \\
              & $\to 0$ & 9.921(1)  & 2.4133(2) \\
              & exact   & 9.921559  & 2.413382  \vspace{0.3em}\\
      $U = 6$ & 0.4000  & 4.66(3)   & 1.490(5)  \\
              & 0.3333  & 4.93(2)   & 1.317(3)  \\
              & 0.2667  & 5.126(8)  & 1.165(1)  \\
              & 0.2000  & 5.29(6)   & 1.036(10) \\
              & 0.1333  & 5.276(7)  & 0.958(1)  \\
              & 0.1000  & 5.31(1)   & 0.924(2)  \\
              & 0.0500  & 5.285(7)  & 0.899(1)  \\
              & $\to 0$ & 5.298(6)  & 0.886(1)  \\
              & exact   & 5.296039  & 0.887600  \\\hline\hline
	\end{tabular}
	\label{tab:ED}
\end{table}
 
\end{document}